\documentclass[12pt,fleqn,a4paper]{article}
\pdfoutput=1
\usepackage{graphicx}
\usepackage[T1]{fontenc}
\usepackage{epsfig,cite}
\usepackage{amssymb}
\usepackage{amsmath}
\usepackage{color}
\usepackage{amstext,alltt,setspace}
\usepackage{amsbsy}
\usepackage{array}
\usepackage{booktabs, makecell}
\usepackage{hyperref}
\usepackage{slashed}
\usepackage{url}
\usepackage{geometry}
\usepackage{tabularx}
\begin{document}
\begin{flushleft}
  \includegraphics[width=.3\linewidth,clip]{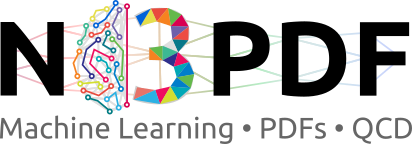}
\end{flushleft}
\vspace{-2.5cm}
\begin{flushright}
  \includegraphics[width=.15\textwidth]{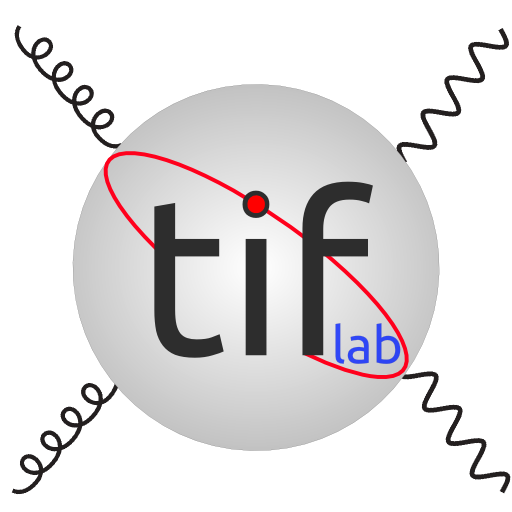}\\
  TIF-UNIMI-2021-1\\
\end{flushright}

\vspace*{.2cm}

\begin{center}
  {\Large \bf{Future tests of parton distributions }}
\end{center}

\vspace*{.7cm}

\begin{center}
  Juan Cruz-Martinez$^a$, Stefano Forte$^a$ and Emanuele R.~Nocera$^b$
  \vspace*{.2cm}

{  \noindent
      {\it
        ~${}^a$Tif Lab, Dipartimento di Fisica, Universit\`a di Milano and\\ 
        INFN, Sezione di Milano,
        Via Celoria 16, I-20133 Milano, Italy\\[0.1cm]

          ~${}^b$The Higgs Centre for Theoretical Physics, University of Edinburgh,\\
  JCMB, KB, Mayfield Rd, Edinburgh EH9 3JZ, Scotland\\}}

      \vspace*{3cm}

      {\bf Abstract}
\end{center}

{\noindent We discuss a test of the generalization power of the
  methodology used in the determination of parton distribution functions (PDFs). The
  ``future test'' checks whether the uncertainty on PDFs, in regions in which they are not constrained by
  current data, are compatible with future data. The test is performed
  by using the current optimized methodology for PDF determination, but
  with a limited  dataset, as available in the past, and by checking
  whether results are compatible within uncertainty with the result
  found using a current more extensive dataset. 
   We use the future test to assess the generalization power of the
   NNPDF4.0 unpolarized PDF and the NNPDFpol1.1 polarized PDF
   methodology. Specifically, we investigate whether 
   the former would predict the rise of the unpolarized proton structure
   function $F_2$ at small $x$ using only pre-HERA data, and whether the latter
   would predict the  so-called ``proton spin crisis'' using only pre-EMC data.
}
\begin{center}

\bigskip
\bigskip
{\it Prepared for the 60th anniversary of the Cracow School of
  Theoretical Physics\\
  to be published in the proceedings}
\end{center}

\pagebreak
\begin{flushright}
  "But how can I use a method to discredit that very method,\\ if the
  method is discreditable?"~\footnote{Jak\.ze jednak u\.zy\'c metody,
    kt\'ora dzi\k eki sobie samej ma zosta\'c podana w
    nies\l aw\k e?
  }\\
  Stanislaw Lem, {\it The futurological Congress}~\cite{LemFC}
  \end{flushright}

\section{Parton distributions in the era of precision}
\label{sec:intro}
Knowledge of parton distribution functions (PDFs), or lack thereof,
is currently one of the dominant sources of uncertainty in the
computation of LHC processes. Over the last several years, as the
precision of PDF determinations has been gradually improving, a major
issue has been that of making sure that this increased precision is matched by
a correspondingly high accuracy. Specifically, the question is whether
uncertainties due to the methodology used in PDF extraction are
estimated accurately. As well known, the problem in determining parton
distributions is lack of knowledge of their underlying functional
form. Because one is extracting a probability distribution for a set
of functions from a discrete set of data points, it is necessary to
make assumptions in order to make the problem solvable. The question
is then how to reliably estimate the uncertainty coming from these
assumptions.

Over the last several years, it has been shown that this problem can
be successfully handled using machine learning techniques, by
essentially viewing it as a pattern recognition problem (see
Ref.~\cite{Forte:2020yip} for a recent review). The basic idea is
that standard tools, such as neural networks, can be
used to infer an underlying pattern, i.e.~the shape of the PDF, without having to assume a functional form. Redundancy
then ensures that no bias is introduced, while quality control tools,
such as cross-validation, ensure that the most general result which is
compatible with the data is obtained, without also reproducing
statistical noise. Uncertainties are obtained as the spread of a
population of equally good best-fits. The reader is referred to more
detailed discussions~\cite{Forte:2010dt,Gao:2017yyd,Ethier:2020way,Forte:2020yip} for a treatment of this machine-learning based
methodology and how it relates to other approaches used for PDF
determination.

This paper deals with the issue of validating PDFs, specifically those
determined using machine learning methods. As we shall see, this
validation problem really consists of two different problems: the first
has to do with the ability of machine learning tools to learn an
unknown underlying law, and the second has to do with their
generalization power. Whereas there are now relatively
well-established tools for the validation of the former, here we
discuss a new tool for the validation of the latter: the ``future
test''. We will first discuss in Section~\ref{sec:MLPDFs} the issue of
PDF validation. In order to make the discussion self-contained, we
briefly review, in Section~\ref{sec:factkin}, the relation between
PDFs and the underlying data, and we  then turn, in
Section~\ref{sec:ctft}, to an explanation of the reason why a new
validation method is needed, specifically in order to address
generalization, and to a presentation of the idea of the future test. In the
remaining two sections, we see the future test at work, by applying it
to
unpolarized (Section~\ref{sec:CTunpol}) and polarized
(Section~\ref{sec:CTpol}) PDFs. Specifically, we discuss the small $x$ rise of
the unpolarized structure function $F_2$ at HERA, and the ``spin
crisis'', related to the first measurements of the polarized structure
function $g_1$. In both cases, we address the issue whether these
then-surprising discoveries could have been predicted, had modern-day
methodology been available.

\section{Validation of PDFs}
\label{sec:MLPDFs}

The original motivation for introducing machine learning tools in PDF
determination~\cite{Forte:2002fg} is to avoid the sources of bias
involved in other methodologies, such as the need to pick a particular
functional form, or an  insufficiently flexible choice of model.
The claim that a certain methodology is unbiased, or less biased,
however, immediately raises the question:
how can we make
sure that results, and specifically uncertainties, are faithful?
The choice of 
using  machine learning tools suggests that validation should be
performed a posteriori, given the lack of direct control on the
underlying model. In searching for an a posteriori validation
methodology  there are then two rather distinct issues, corresponding to
the kinematic regions we consider.

\subsection{Factorization and kinematic regions}
\label{sec:factkin}

In order to understand the different issues related to PDF validation
it is necessary to briefly recall the way PDFs are related to the
observable cross sections that are used in order to determine
them. Consider specifically a hadronic collision.
A typical 
cross section $\sigma(Q^2,\tau,\{k\})$ is expressed in terms of PDFs as
\begin{equation}\label{eq:fact}
   \sigma(Q^2,\tau,\{k\})=\sum_{ij}\int_\tau^1\frac{dz}{z} {\cal L}_{ij}
   (z, Q^2) \hat \sigma_{ij}\left(\frac{\tau}{z},\alpha_s(Q^2),\{k\}\right),
\end{equation}
where $\hat \sigma_{ij}$ is the (``partonic'') cross section for producing the
desired final state with incoming partons $i$, $j$,
$Q^2$ is a hard scale (for example, for $Z$ production, the $Z$
mass $Q^2=M^2_Z$), $\tau$ is a dimensionless scaling variable (for example, for the
total inclusive $Z$ production cross section, the ratio
$\tau=M_Z^2/s$, with $s$ the center-of-mass energy), and $\{k\}$
denotes the set of other kinematic variables that the (generally
differential) cross section $\sigma$ may depend upon (for example, the
rapidity of the dilepton pair into which the $Z$ decays). The
dependence on the PDFs is contained in the luminosity
\begin{equation}\label{eq:lumi}
    {\cal L}_{ij}(z,Q^2)=\int_x^1 \frac{dx}{x} f_i^{h_1}(x,Q^2)
   f_j^{h_2}\left(\frac{z}{x},Q^2\right),
\end{equation}
where $f^{h_k}_i(x,Q^2)$ is the PDF for extracting a parton of species
$i$ from hadron $h_k$, carrying a fraction $x$ of its momentum at
scale $Q^2$.
\begin{figure}[t]
\begin{center}
  \includegraphics[width=0.49\textwidth]{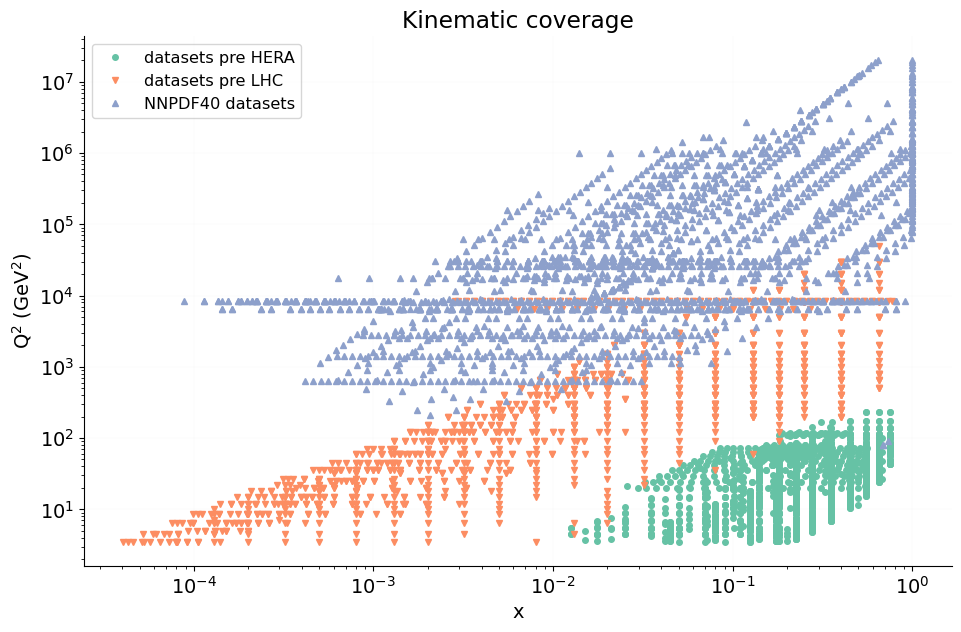}
  \includegraphics[width=0.49\textwidth]{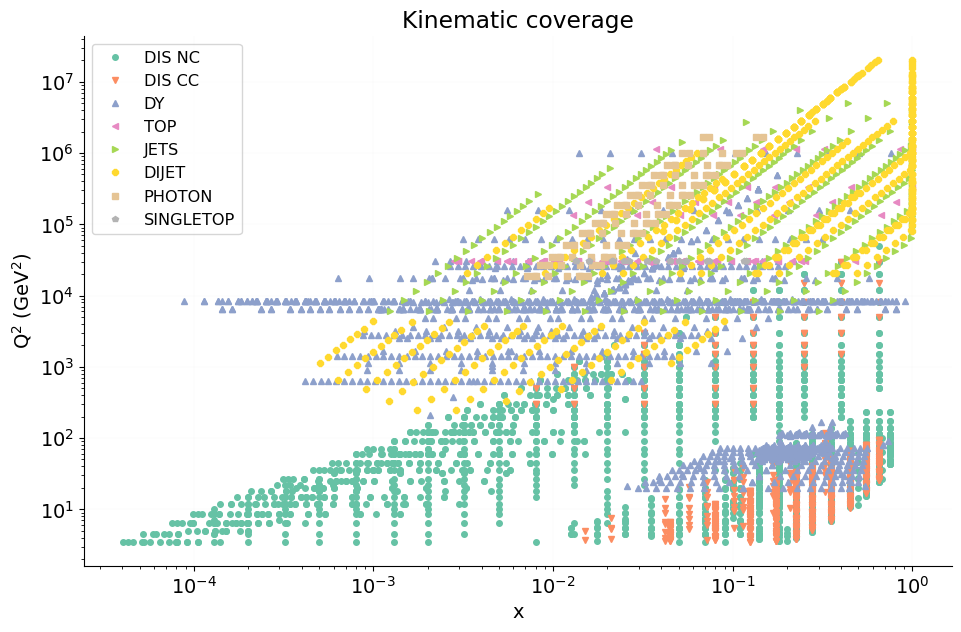}
  \caption{\small Scatter plot for the datapoints entering the
    NNPDF4.0 PDF
    determination and future tests in the unpolarized case, listed in Table~\ref{tab:dat}. The data
    are shown color-coded by future test grouping (left), and the full
    dataset is also shown color-coded 
    by physics process (right).
   }
    \label{fig:scatter}
\end{center}
\end{figure}

Now, the structure of Eq.~(\ref{eq:lumi}) trivially implies that
only PDFs in a certain range are probed by any given physical process:
specifically, the PDF for $x<\tau$ completely decouples from an
observable measured at $\tau$. Furthermore, the dominant contribution
to any given process typically comes from a restricted kinematic
region: in the simplest case, from a single  kinematic point. For
example, if the $Z$ rapidity distribution is evaluated at leading
order, the partonic cross section $\hat\sigma$ is proportional to a
two-dimensional Dirac delta which removes the convolution integrals from
Eq.~(\ref{eq:lumi}), and only PDFs evaluated for the fixed values
$x_1=\tau\exp\left(\frac{1}{2}\eta\right)$, $x_2=\tau\exp-\left(\frac{1}{2}\eta\right)$, $Q^2=M^2_Z$ contribute, with $\eta$
the rapidity of the $Z$ boson. Even when the kinematics is
completely fixed, it is only fixed  at leading perturbative order;
at higher perturbative orders the cross section is no longer a Dirac delta,
and all values of $x$ which are larger than that fixed by
leading-order kinematics are accessible, with the smaller $x$ region remaining
inaccessible. This is easy to understand physically: at higher
perturbative order a parton can always radiate other partons, so all
partons that carry a 
momentum fraction of their parent hadron which is larger than the minimal
value needed in order to produce the desired final state can
contribute. However, they must at least carry the minimal required
momentum fraction. Therefore, for any given process, there is a
minimal value of $x$, though in principle no maximal value.

In Fig.~\ref{fig:scatter} we show a scatter plot of data used for a current
determination of unpolarized  PDFs: each point corresponds
to the value of $x$ and $Q^2$ at which the PDF contributes to each
datapoint, determined using leading-order kinematics, whenever the
kinematics of the leading-order process completely fixes them, and the
minimum value of $x$, whenever it does not fix them completely (such
as for single-inclusive jets). Note that, as already mentioned, the region of
$x$ smaller than the smallest $x$ datapoints of Fig.~\ref{fig:scatter}
is completely inaccessible. The larger $x$ region is in principle
accessible at higher orders, though in practice this is hardly the
case for sufficiently large $x$ (say $x\gtrsim0.5$). Indeed, PDFs vanish at $x=1$ and drop quite fast as $x$
increases (typically as a power of $(1-x)$) so, for many processes,
not only the contribution from
larger $x$ arises at higher
perturbative orders, but also, it rapidly becomes much smaller than
that from the smallest value of $x$, thus being in practice unobservable.

Inspection of Fig.~\ref{fig:scatter}  immediately shows that 
the PDF determination, and its validation, entails two rather distinct
issues, related to the fact that data points are localized in a
well-defined, connected kinematic region --- the ``data region'', henceforth.
On the one hand, the problem of
determining PDFs in the region probed by the data remains
mathematically ill-posed, because one is determining a function from a piece of
discrete information. Yet, because of the convolution
integral, PDFs are not probed at a single point, but rather in a
region, and also, even in the polarized case, where data are more
scarce, datapoints are typically dense in the experimentally
accessible kinematic region, because experimental collaborations strive
to measure physical observables with as fine a binning as
possible. Consequently, whereas lack of knowledge of the PDF
functional form entails that a population of equally likely best-fits
to any given set of data is possible, assumptions of smoothness, which
are physically justified, and
built in standard machine-learning interpolants (such as neural
networks) ensure that a reasonably narrow set of results follow.

On the other hand, in the small and large $x$
regions which are not probed by the data ---
the ``extrapolation region'', henceforth --- 
the uncertainty is a priori infinite, unless
one introduces some assumptions allowing for generalization of the
behavior observed in the region covered by the data. Infinite
uncertainties would imply the impossibility of predicting any hadronic
process, such as Higgs production. Indeed, the bulk of the PDF
information which is necessary to predict a given LHC
process can be estimated using suitable tools~\cite{Carrazza:2016htc},
but a certain amount of extrapolation is always needed.
The only way to avoid this
unpalatable conclusion is to introduce some assumption --- motivated by
physical considerations, or perhaps based on extrapolating, or
generalizing a behavior seen in the data region.

\subsection{Validation methods: closure tests and future tests}
\label{sec:ctft}

We can now discuss validation in the data and extrapolation region,
and why they differ. 
The validation of PDF uncertainties in the data region has been
addressed in Ref.~\cite{Ball:2014uwa} (see also
Refs.~\cite{Forte:2020yip,NNPDF40,LDDMW} for an update): it can be performed
using the closure test methodology. The idea is to assume a particular
``true'' underlying form of the PDFs, generate data based on this
assumption, with some statistical distribution corresponding to a
particular covariance matrix (typically, the same as that of the true
experimental data) and then perform the PDF determination on this data. The
faithfulness of the uncertainties can be checked by comparing the
distribution of answers when the procedure is repeated several times
with the true value, which is now known. Because both the experimental
uncertainties and the theory are now completely correct by
construction --- they are used in the data generation --- the
procedure tests the accuracy and reliability of the methodology.

The validation of the methodology in the extrapolation region is an
open issue: it is  the
problem that we would like to address here. There is a fundamental reason
why this cannot be approached in the same way as the validation in the
data region, so in particular a closure test is not advisable. This is
related to the fact that
the current methodology, if correctly implemented, incorporates the
constraints which are present (even if sometimes not explicitly known)
in the current data.
In order to understand this, consider a simple
example. It is known that PDFs satisfy sum rules (see
e.g. Ref.~\cite{Forte:2010dt}) : for example the momentum sum rule
\begin{equation}\label{eq:msrule}
  \sum_i\int_0^1\, dx xf_i(x,Q^2)=1
  \end{equation}
 where the sum runs over all parton
 species, which holds for all values of $Q^2$. Equation~(\ref{eq:msrule}) expresses the fact
that the total momentum carried by all partons adds up to the momentum
of their parent hadron. Because this is a known property of QCD, it
is usually imposed as a hard-wired constraint; however it can be
checked that if it is not imposed it is actually reproduced by the
PDFs extracted from the data~\cite{Ball:2011uy}.

Now, the machine learning
methodology for PDF determination is actually optimized on existing
data. This optimization was done by trial and error in the past, and it is
currently performed, at least in part, using automatic
hyperoptimization
techniques~\cite{Carrazza:2019mzf,Forte:2020yip}. Clearly, this
optimization is performed in the space of possible solutions, which is
a subspace of all possible PDFs, incorporating both known constraints, such as the
momentum sum rule Eq.~(\ref{eq:msrule}), but also  possible
constraints that we do not know of.

Assume now for the sake of argument that
we did not know the momentum sum rule. A methodology optimized on
current data would work well on data that respect it. In order to
closure-test the methodology we have to assume an underlying
functional form. Typically, we would assume that a true PDF is one PDF
as determined now, and so even if we did not know the momentum sum
rule we would generate data that respect it to good approximation, and
use them to closure-test a methodology that unknown to us
incorporates this constraint. So the test would be meaningful.

But assume now that we generated data {\it outside} the currently
measured region. We would then have to make some assumption on the
underlying PDF there, and this might not incorporate true
constraints that hold there, or it might actually include some
constraints that in actual fact do not really hold. In our example, if
we did not know the momentum sum rule, we might generate pseudodata
that violate it. A validation based on
this procedure would then be unreliable because there is no way to know whether 
there is a mismatch between constraints in the generated data and the
methodology. In other words, closure testing is only effective if the
generated data carry an amount of information that is smaller, or at
most comparable, to real data. If the generated data carry more
information--- as it must be the case if closure testing the
extrapolation region --- there is no way to know whether a bias in
the methodology will go undetected, or whether the test would force a
bias upon the methodology itself.

The idea of the future test is to verify whether a methodology
optimized on current data generalizes correctly the features contained
in a subset of this data. So, for example, a future test of the
methodology used for the NNPDF3.1 determination of
PDFs~\cite{Ball:2017nwa} consists of using the same methodology to
extract PDFs from, say, only the deep-inelastic scattering (DIS) subset of
the NNPDF3.1 dataset. 
The future test then amounts to
comparing these PDFs to the full NNPDF3.1 dataset. Indeed, DIS data cannot constrain the large
$x$ gluon PDF, which is instead well determined by LHC data in the
NNPDF3.1 dataset~\cite{Nocera:2017zge}.  So, if one compares these data
to PDFs determined from DIS only, one is testing whether the methodology
manages to generalize the gluon PDF outside the data region.

The simplest
way to perform the test is to view the computation of all data not
included in the PDF fit as a prediction, and compare this prediction,
with its PDF uncertainty, with the actual value. The future test is
then successful if the $\chi^2$ per data point is of order one, because
this means that the PDF uncertainties have been estimated faithfully
in the extrapolation region. More quantitatively, a criterion for the
future test to be successful can be spelled out as
\begin{equation}
\label{eq:criterion}
  |{\chi^2}^{\rm PDF}_{\rm not~fitted}-\chi^2_{\rm fitted}|\ll
|{\chi^2}^{\rm PDF}_{\rm not~fitted}-\chi^2_{\rm not~fitted}|,
\end{equation}
where we denote by ${\chi^2}^{\rm PDF}$ and ${\chi^2}$ respectively
the $\chi^2$ per datapoint with and without PDF uncertainties
included. In other words, the test is successful if the inclusion of
the PDF uncertainties accounts for the missing information which leads
the predictions for  data which are not fitted to deviate from the
measured values.

The test can be thought of as the comparison of a prediction of future
data with the future data themselves - hence the name. 
However, this is a manner of
speaking: it is important to understand what the future test does and does not
test. The future test does not test whether current data could have
been predicted in the past, because the future test tests the {\it current}
methodology. Indeed, the methodology that we are future testing is
optimized to the current dataset: our goal is to test the
generalization power of our current best methodology. To this purpose,
we test the capability of this methodology of extracting features
of the full dataset from a subset, i.e. its generalization
power. Also, the future test relies on all the information which is
available on the full present-day dataset, both from an experimental
and a theoretical point of view. So, for example, we now know that
some data must be treated at the highest perturbative accuracy which
is currently available, and these data are therefore excluded from PDF
determinations which are now performed at a lower perturbative order
--- but higher order corrections may have been unavailable
when these data were originally
published. 

In
this sense, the future test is weaker than a closure test. Indeed, a
successful future test is necessary but not sufficient in order to
guarantee that the extrapolation of current PDFs outside their data
region is reliable. It could be that a reliable extrapolation
requires information that is currently inaccessible: however, this is
the best we can do now. This is akin to the situation that arises when
estimating uncertainties related to missing higher order terms in a
perturbative computation. The best we can do is to base the
estimate on the known terms, but it cannot be excluded that some new
piece of information is needed, which only arises at some higher
perturbative order, in which case the estimate is unreliable.
Even so, the future test poses  stringent requirements on PDFs, as we
will show in explicit examples in the next sections.

\begin{table}[p]
\renewcommand*{\arraystretch}{1.60}
\scriptsize
\centering
\begin{tabularx}{.49\textwidth}{|Xcr|}
\toprule
{\bf pre-HERA}
    & Ref. 
    & $N_{\rm{dat}}$ \\
\midrule
NMC $d/p$
    & \cite{Arneodo:1996kd}
    &  121 \\
NMC $p$
    & \cite{Arneodo:1996qe}
    &  204 \\
SLAC $p$
    & \cite{Whitlow:1991uw}
    &   33 \\
SLAC $d$
    & \cite{Whitlow:1991uw}
    &   34 \\
BCDMS $p$
    & \cite{Benvenuti:1989rh}
    &  333 \\
BCDMS $d$
    & \cite{Benvenuti:1989fm}
    &  248 \\
CHORUS $\sigma_{CC}^{\nu}$
    & \cite{Onengut:2005kv}
    &  416 \\
CHORUS $\sigma_{CC}^{\bar{\nu}}$
    & \cite{Onengut:2005kv}
    &  416 \\
NuTeV $\sigma_{c}^{\nu}$
    & \cite{Goncharov:2001qe}
    &   39 \\
NuTeV $\sigma_{c}^{\bar{\nu}}$
    & \cite{Goncharov:2001qe}
    &   37 \\
DYE 866 $\sigma^d_{\rm DY}/\sigma^p_{\rm DY}$
    & \cite{Towell:2001nh}
    &   15 \\
DY E886 $\sigma^p_{\rm DY}$
    & \cite{Webb:2003ps}
    &   89 \\
DY E605 $\sigma^p_{\rm DY}$
    & \cite{Moreno:1990sf}
    &   85 \\
\midrule
Total pre-HERA
    &
    & 2070 \\
\midrule
{\bf pre-LHC}
    & Ref. 
    & $N_{\rm{dat}}$ \\ 
\midrule
HERA I+II inclusive NC $e^-p$
    & \cite{Abramowicz:2015mha}
    &  159 \\
HERA I+II inclusive NC $e^+p$ 460 GeV
    & \cite{Abramowicz:2015mha}
    &  204 \\
HERA I+II inclusive NC $e^+p$ 575 GeV
    & \cite{Abramowicz:2015mha}
    &  254 \\
HERA I+II inclusive NC $e^+p$ 820 GeV
    & \cite{Abramowicz:2015mha}
    &   70 \\
HERA I+II inclusive NC $e^+p$ 920 GeV
    & \cite{Abramowicz:2015mha}
    &  377 \\
HERA I+II inclusive CC $e^-p$
    & \cite{Abramowicz:2015mha}
    &   42 \\
HERA I+II inclusive CC $e^+p$
    & \cite{Abramowicz:2015mha}
    &   39 \\
HERA comb. $\sigma_{c\bar c}^{\rm red}$
    & \cite{H1:2018flt}
    &   37 \\
HERA comb. $\sigma_{b\bar b}^{\rm red}$
    & \cite{H1:2018flt}
    &   26 \\
CDF $Z$ rapidity
    & \cite{Aaltonen:2008eq}
    &   28 \\
D0 $Z$ rapidity
    & \cite{Abazov:2013rja}
    &   28 \\
D0 $W\to \mu\nu$ asymmetry
    & \cite{Abazov:2013rja}
    &    9 \\
\midrule
Total pre-LHC
    &
    & 1273 \\
\midrule
{\bf NNPDF4.0}                               
    & Ref. 
    & $N_{\rm{dat}}$ \\
\midrule
ATLAS $W,Z$ 7 TeV
    & \cite{Aad:2011dm}
    & 30 \\
ATLAS HM DY 7 TeV
    & \cite{Aad:2013iua}
    &  5 \\
ATLAS low-mass DY 7 TeV
    & \cite{Aad:2014qja}
    & 6  \\
ATLAS $W,Z$ 7 TeV central selection
    & \cite{Aaboud:2016btc}
    & 46 \\
ATLAS $W,Z$ 7 TeV forward selection
    & \cite{Aaboud:2016btc}
    & 15 \\
ATLAS DY 2D 8 TeV
    & \cite{Aad:2016zzw}
    & 48 \\
ATLAS $W,Z$ inclusive 13 TeV
    & \cite{Aad:2016naf}
    & 3  \\
ATLAS $W^+$+jet 8 TeV
    & \cite{Aaboud:2017soa}
    & 16 \\
\bottomrule
\end{tabularx}
\hfill
\renewcommand*{\arraystretch}{1.625}
\begin{tabularx}{.49\textwidth}{|Xcr|}
\toprule
{\bf NNPDF4.0 (cont.'d)}
    & Ref. 
    & $N_{\rm{dat}}$ \\  
\midrule
ATLAS $W^-$+jet 8 TeV
    & \cite{Aaboud:2017soa}
    & 16 \\
ATLAS $Z$ $p_T$ 8 TeV $(p_T^{ll},M_{ll})$
    & \cite{Aad:2015auj}
    & 44 \\
ATLAS $Z$ $p_T$ 8 TeV $(p_T^{ll},y_{ll})$
    & \cite{Aad:2015auj}
    & 48 \\
ATLAS $\sigma_{tt}^{\rm tot}$ 7, 8, 13 TeV
    & \cite{Aad:2014kva,Aad:2020tmz}
    & 3  \\
ATLAS $t\bar{t}$ $y_t$ normalized 8 TeV
    & \cite{Aad:2015mbv}
    & 4  \\
ATLAS $t\bar{t}$ $y_{t\bar{t}}$ normalized 8 TeV
    & \cite{Aad:2015mbv}
    & 4  \\
ATLAS $t\bar{t}$ normalized $|y_t|$ dilepton 8 TeV
    & \cite{Aaboud:2016iot}
    & 5  \\
ATLAS jets 8 TeV, R=0.6
    & \cite{Aaboud:2017dvo}
    & 171 \\
ATLAS dijets 7 TeV, R=0.6
    & \cite{Aad:2013tea}
    & 90 \\
ATLAS direct photon production 13 TeV
    & \cite{Aaboud:2017cbm}
    & 53 \\
ATLAS single top $R_{t}$ 7 TeV
    & \cite{Aad:2014fwa}
    & 1  \\
ATLAS single top $R_{t}$ 13 TeV
    & \cite{Aaboud:2016ymp}
    & 1  \\
ATLAS single top $y_t$ norm. 7, 8 TeV
    & \cite{Aad:2014fwa,Aaboud:2017pdi}
    & 6  \\
ATLAS single antitop $y$ norm. 7, 8 TeV
    & \cite{Aad:2014fwa,Aaboud:2017pdi}
    & 6  \\
CMS $W$ electron asymmetry 7 TeV
    & \cite{Chatrchyan:2012xt}
    & 11 \\
CMS $W$ muon asymmetry 7 TeV
    & \cite{Chatrchyan:2013mza}
    & 11 \\
CMS Drell-Yan 2D 7 TeV
    & \cite{Chatrchyan:2013tia}
    & 110 \\
CMS $W$ rapidity 8 TeV
    & \cite{Khachatryan:2016pev}
    & 22 \\
CMS $Z$ $p_T$ $(p_T^{ll},y_{ll})$ 8 TeV
    & \cite{Khachatryan:2015oaa}
    & 28 \\
CMS dijets 7 TeV
    & \cite{Chatrchyan:2012bja}
    & 54 \\
CMS 3D dijets 8 TeV
    & \cite{Sirunyan:2017skj}
    & 122 \\
CMS $\sigma_{tt}^{\rm tot}$ 7, 8, 13 TeV
    & \cite{Spannagel:2016cqt,Khachatryan:2015uqb}
    & 3  \\
CMS $t\bar{t}$ rapidity $y_{t\bar{t}}$ $\ell$+jet 8 TeV
    & \cite{Khachatryan:2015oqa}
    & 9  \\
CMS $\sigma_{tt}^{\rm tot}$ 5 TeV
    & \cite{Sirunyan:2017ule}
    & 1  \\
CMS $t\bar{t}$ 2D $2\ell$ $(m_{t\bar{t}},y_{t})$ 8 TeV
    & \cite{Sirunyan:2017azo}
    & 16 \\
CMS $t\bar{t}$ absolute $y_t$ $\ell$+jet 13 TeV
    & \cite{Sirunyan:2018wem}
    & 10 \\
CMS $t\bar{t}$ absolute $|y_t|$ $2\ell$ 13 TeV
    & \cite{Sirunyan:2018ucr}
    & 11 \\
CMS single top $\sigma_{t}+\sigma_{\bar{t}}$ 7 TeV
    & \cite{Chatrchyan:2012ep}
    & 1  \\
CMS single top $R_{t}$ 8 TeV
    & \cite{Khachatryan:2014iya}
    & 1  \\
CMS single top $R_{t}$ 13 TeV
    & \cite{Sirunyan:2016cdg}
    & 1  \\
LHCb $Z$ 940 pb
    & \cite{Aaij:2012mda}
    & 9  \\
LHCb $Z\to ee$ 2 fb
    & \cite{Aaij:2015vua}
    & 17 \\
LHCb $W,Z \to \mu$ 7 TeV
    & \cite{Aaij:2014wba}
    & 29 \\
LHCb $W,Z \to \mu$ 8 TeV
    & \cite{Aaij:2015zlq}
    & 30 \\
LHCb $Z\to \mu\mu$ 13 TeV
    & \cite{Aaij:2016mgv}
    & 16 \\
LHCb $Z\to ee$ 13 TeV
    & \cite{Aaij:2016mgv}
    & 15 \\
\midrule
Total NNPDF4.0
    &
    & 1148   \\
\midrule
Grand total
    & 
    & 4491   \\
\bottomrule
\end{tabularx}

\vspace{0.1cm}
\caption{\small The pre-HERA, pre-LHC and NNPDF4.0 datasets: for each
  experiment the reference to the original publication and the
  number of datapoints are given. Pre-HERA PDFs are
  fitted to pre-HERA data, pre-LHC PDFs are fitted to the pre-HERA
  and pre-LHC data, and NNPDF4.0 PDFs are fitted to the union of the
  three datasets.}
\label{tab:dat}
\end{table}

\section{Unpolarized PDFs and the rise of structure functions}
\label{sec:CTunpol}

We start by presenting a future test of the NNPDF4.0 methodology. This
is a methodology based, for the first time,  on automatic
hyperoptimization~\cite{Carrazza:2019mzf} (see \cite{Forte:2020yip}
for a review). PDFs constructed with this methodology are currently in
a testing phase and will be released soon~\cite{NNPDF40}. This
methodology is the state of the art in PDF determination based on
machine learning, and it is used for a PDF determination based on a
dataset whose size exceeds any previous determination, specifically
including a large number of LHC Run~II data. Consequently, PDF
uncertainties are smaller than in any previous PDF determination, and
validation issues are particularly important. The validation of NNPDF4.0
PDFs in the data region, performed using an updated version~\cite{Forte:2020yip} of
the closure test methodology of Ref.~\cite{Ball:2014uwa}, will be
discussed in Ref.~\cite{NNPDF40}. 

We choose to use the most recent NNPDF4.0 methodology as a first
illustration of the future test because PDF uncertainties found with
this methodology are relatively smaller, and thus the issue of
accurately estimating PDF uncertainties in extrapolation is
particularly serious. However, the methodology that we will discuss in
Section~\ref{sec:CTpol} in the polarized case is essentially the same
as the unpolarized NNPDF3.1 methodology: hence this methodology will
also be implicitly future-tested there. A direct comparison between
future tests of subsequent NNPDF methodologies is an interesting topic
that will be left for future studies~\cite{NNPDF40}. A relevant
observation, in this respect, is that all published NNPDF PDF sets
turned out to be forward-backward compatible, in the sense that more
recent PDFs have smaller uncertainties, but are compatible within
uncertainties with previously published ones. 

\subsection{Gluons and HERA, quarks and the LHC}
\label{sec:gHqL}

We performed future tests of the NNPDF4.0 methodology based on two datasets of
increasing size: a pre-HERA dataset, which only includes fixed-target
DIS and Drell-Yan production data; and a dataset obtained by combining
this with a  pre-LHC dataset, which
also includes HERA DIS data and Tevatron collider $W$, $Z$ production
and single-inclusive jet data. The NNPDF4.0 PDFs are determined by
adding to these also an NNPDF4.0 dataset, which includes LHC data
for $W$ and $Z$ production, single-inclusive jets and dijets, $Z$
transverse momentum distributions, single top and top pair production, and
prompt photon production. Note that PDFs are thus determined from
strictly hierarchical datasets: pre-HERA PDFs are determined from the
pre-HERA data, pre-LHC PDFs are determined combining pre-HERA and
pre-LHC data, and NNPDF4.0 PDFs are determined combining pre-HERA,
pre-LHC, and NNPDF4.0 data.
Although, as explained in
Section~\ref{sec:ctft} the historical nature of these datasets is
somewhat besides the point, since the future test is a test of 
  current PDF methodology, it might be useful to think
about these datasets roughly as those on  which PDF determinations
were respectively based circa 1993, such as the
CTEQ1~\cite{Botts:1992yi} or MRS~\cite{Martin:1992as} PDFs, for the
pre-HERA PDFs, and circa 2010, such as NNPDF2.1~\cite{Ball:2010de} for
the pre-LHC PDFs.

The three datasets are listed in Table~\ref{tab:dat} and displayed in Fig.~\ref{fig:scatter}, where the
breakdown of the full dataset into single processes is also shown.
The choice of these two future test datasets is motivated
by aspects having to do both with physics and methodology. In terms of
physics, the pre-LHC dataset only imposes very weak constraints on the
flavor separation of quarks and antiquarks, and also on the large-$x$
gluon. In the pre-LHC dataset, the small $x$ gluon is also
unconstrained, so in fact the gluon is largely unconstrained,
except in a very small region around $x\sim 0.1$. In terms of
methodology, a pre-LHC fit tests whether relatively small PDF
uncertainties (of order of say 5-10\%) are compatible with later more
precise data: i.e., it tests the ability of the methodology to perform
a near extrapolation. On the contrary, the pre-HERA fit tests
whether the methodology can capture broad features of the results even
in the presence of a minimal amount of information: i.e. it tests the
far extrapolation.

A particularly intriguing aspect of the pre-HERA future test in this
respect is the small $x$ behavior of the gluon PDF, and correspondingly
of the $F_2$ proton structure function. Indeed, when this structure
function was first measured at HERA the observed rise of the structure
function at small $x$ (see e.g.~\cite{DeRoeck:1995mt}) came as a
surprise, and indeed standard pre-HERA PDF sets displayed a wide
variety of small-$x$ gluon behaviors, with  a flat gluon taken as a
baseline option in the absence
of new physics effects~\cite{Martin:1992as}. A rather steeper rise of the
gluon at small $x$ is a common feature of all post-HERA PDF sets
(see~\cite{Tung:2004rw} for a comparative discussion). It is thus
interesting to wonder how a contemporary methodology, presented with
pre-HERA data, would behave in this respect.

\subsection{Future testing NNPDF4.0}
\label{sec:40}

\begin{table}[t]
\renewcommand*{\arraystretch}{1.60}
\scriptsize
\centering
\begin{tabularx}{\textwidth}{|X|r|ccc||ccc|}
\toprule
                 & $n_{\rm dat}$ &   NNPDF4.0 & pre-LHC & pre-HERA & NNPDF4.0 & pre-HERA & pre-LHC\\
\midrule
pre-HERA dataset & 2070        & 1.09       & 1.01    & 0.90     & (1.02)   & (0.95)   & (0.86) \\
pre-LHC dataset  & 1273        & 1.21       & 1.20    & {\it 23.1}
& (1.18)   & (1.17)   &  {\bf 1.22}  \\
NNPDF4.0 dataset & 1148        & 1.29       & {\it 3.30}    & {\it
  23.1}     & (1.23)   &  {\bf 1.30}    &  {\bf 1.38}  \\
\midrule
Total dataset    & 4491        & 1.17       & 1.65    & 12.9     &  1.12    &  1.10    &  1.10   \\
\bottomrule
\end{tabularx}

\vspace{0.3cm}
\caption{\small The $\chi^2$ per datapoint for the NNPDF4.0 PDF determination
and its pre-HERA and pre-LHC future tests. Values are shown for the
three datasets displayed in Fig.~\ref{fig:scatter}. In the left table
all values are computed using only the experimental covariance matrix,
while in the right table all values are computed by also including PDF
uncertainties. Values with PDF uncertainties included computed for data
included in the fit, shown in parenthesis, do not have a strict
statistical meaning (see text). All numbers in italic (without PDF
uncertainty included) and in boldface (with PDF uncertainty included)
are predictions.}
\label{tab:chi2unp}
\end{table}

We now turn to the results of the future test of the NNPDF4.0 methodology.
 The reader is referred
to the forthcoming Ref.~\cite{NNPDF40} for full details on the
NNPDF4.0 PDF
  determination theory and methodology; here, we provide  some basic
  information. The machine-learning methodology which is being future-tested has been
  constructed through a hyperoptimization
  procedure~\cite{Carrazza:2019mzf} and its general features are
  reviewed in Ref.~\cite{Forte:2020yip}. It uses the same approach,
  based on a Monte Carlo representation of PDF uncertainties, and
  neural networks as underlying interpolants as previous NNPDF
  determinations, such as the most recent published
  NNPDF3.1~\cite{Ball:2017nwa}. It differs from the previous NNPDF
  methodology in the architecture of neural networks and minimization
  algorithm, and more importantly, due to the fact that architecture
  and algorithms have been selected through a hyperoptimization
  procedure tuned using $K$-foldings,
  see~\cite{Forte:2020yip}. Preprocessing is treated according to the
  most recent NNPDF methodology~\cite{Ball:2015oha}.

  The figure of merit which is
  being minimized is the same as in previous  NNPDF determinations, and
  specifically it includes all available information on statistical
  and systematic experimental uncertainties and their correlations,
  through the $t_0$ method~\cite{Ball:2009qv,Ball:2012wy}.
  NNLO QCD theory is used throughout, with charm quark mass effects
  included through the FONLL method~\cite{Forte:2010ta} and a
  parametrized charm
  PDF~\cite{Ball:2015tna,Ball:2015dpa}. Uncertainties due to deuterium
  and heavier nuclear targets are
  included~\cite{Ball:2018twp,AbdulKhalek:2019mzd,Ball:2020xqw} as a theory 
  uncertainty contribution to the covariance
  matrix~\cite{AbdulKhalek:2019ihb}.

  An identical methodology is used to perform the future tests.
    The
  NNPDF methodology calls for the $t_0$ covariance matrix~\cite{Ball:2009qv} and the
  preprocessing exponents 
  to be determined iteratively and
  self-consistently: this has been done for each of the two future
  tests. Results are collected in Table~\ref{tab:chi2unp}, where we
  show $\chi^2$ values for the three PDF sets: namely, the baseline NNPDF4.0,
  the pre-HERA and the pre-LHC future tests. Values are shown for all data,
  both included and not included in each fit. We show both standard
$\chi^2$ values
computed using the experimental covariance matrix, and values computed
also including PDF uncertainties. Note that the
experimental covariance matrix is not the same as the $t_0$
  covariance matrix used for minimization in order to obtain unbiased
  results, but just the  standard covariance matrix as published by
  experimental collaborations (see Ref.~\cite{Ball:2012wy} for a
  comparative discussion). The $\chi^2$ computed using it measures the goodness of fit
  provided by the PDFs fitted to data.

  In the $\chi^2$ including PDF uncertainties, these 
  are accounted for by adding to the experimental covariance
  matrix the PDF uncertainty covariance matrix, in turn computed as
  the covariance over the NNPDF replica
  set~\cite{Demartin:2010er}. The total covariance matrix is then
  inverted and used in the standard definition of the $\chi^2$.
  Addition of two contributions to the covariance matrix is justified
  if they correspond to uncorrelated
  uncertainties~\cite{AbdulKhalek:2019ihb}. This is surely the case
  when considering the PDF uncertainty and the uncertainty on new data
  not used for the determination of those PDF. 
  Consequently, the $\chi^2$ values shown in Table~\ref{tab:chi2unp} with PDF
  uncertainties included,  for the
  future test PDFs when compared to data not used to fit, indicate
  whether the PDF uncertainty estimate on these future test PDFs 
  is reliable: it  tests the generalization power of the
  methodology. 

  Note that, of course, the PDF uncertainty is not uncorrelated to the uncertainty on data
  used for the determination of the same PDFs, since the latter
  propagates into the former.  Consequently
  $\chi^2$ values computed
  including PDF uncertainties for datasets included in any PDF
  determination, shown in parenthesis in Table~\ref{tab:chi2unp}, are only indicative
  and do not have a strict statistical meaning.

It is important to stress that both the experimental covariance
matrix, and the PDF covariance matrix, consistently include
correlations. In fact, these are often quite large, and $\chi^2$
values found not including correlations would be substantially
different. For the experimental covariance matrix, this is because
for many LHC data uncorrelated statistical uncertainties are
very small: sometimes, for example for $Z$ production data, at the
per mille level. Dominant uncertainties are then systematic, and these
are typically highly or fully correlated: an example is the luminosity
uncertainty, which is common to all data in a given experiment. For the
PDF covariance matrix this is due to the fact that many data points
are kinematically very close to each other, and of course PDF values
and uncertainties on neighboring values of $x$ and $Q^2$ are highly correlated.
  
\begin{figure}[t]
\begin{center}
  \includegraphics[width=0.49\textwidth]{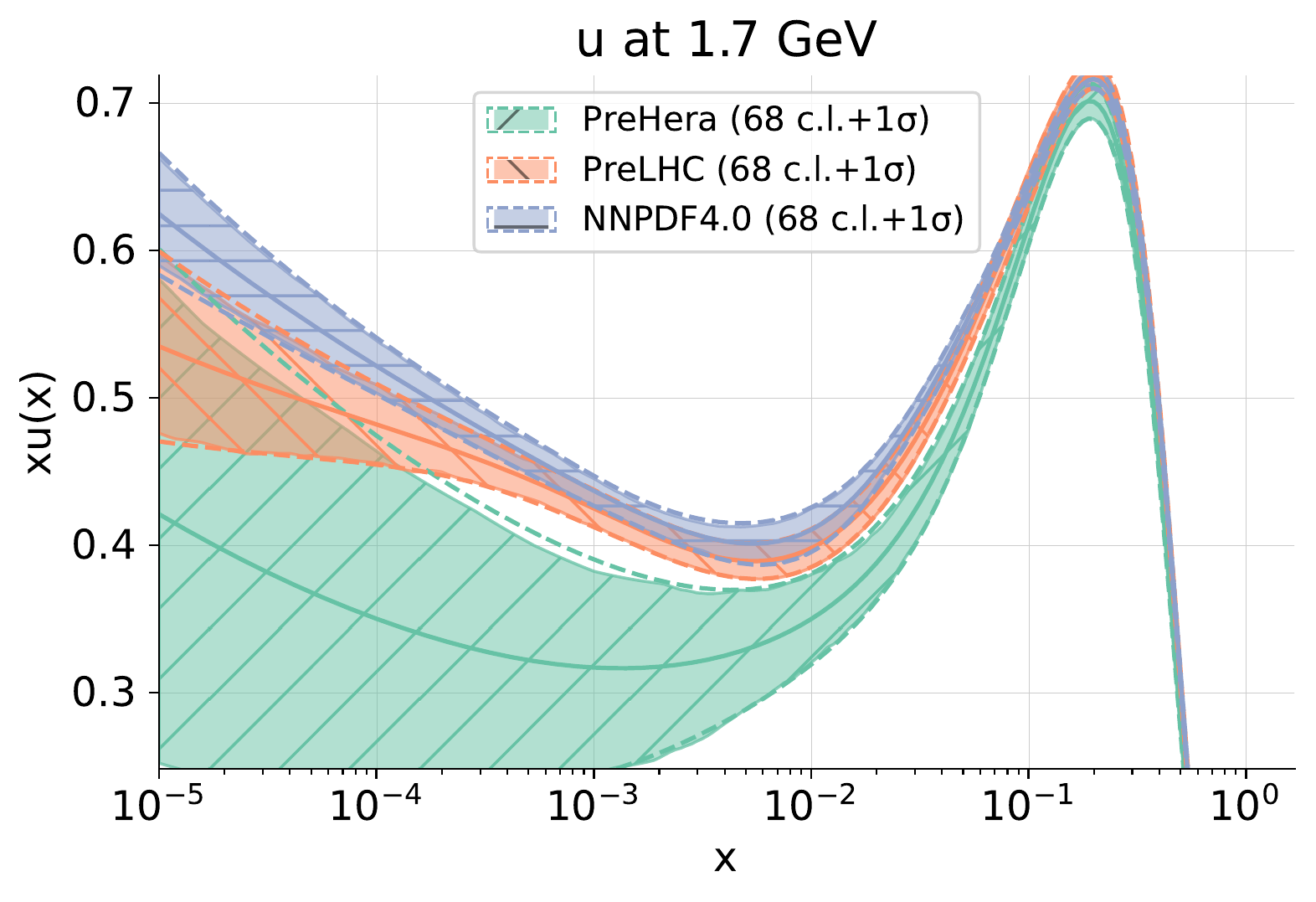}
  \includegraphics[width=0.49\textwidth]{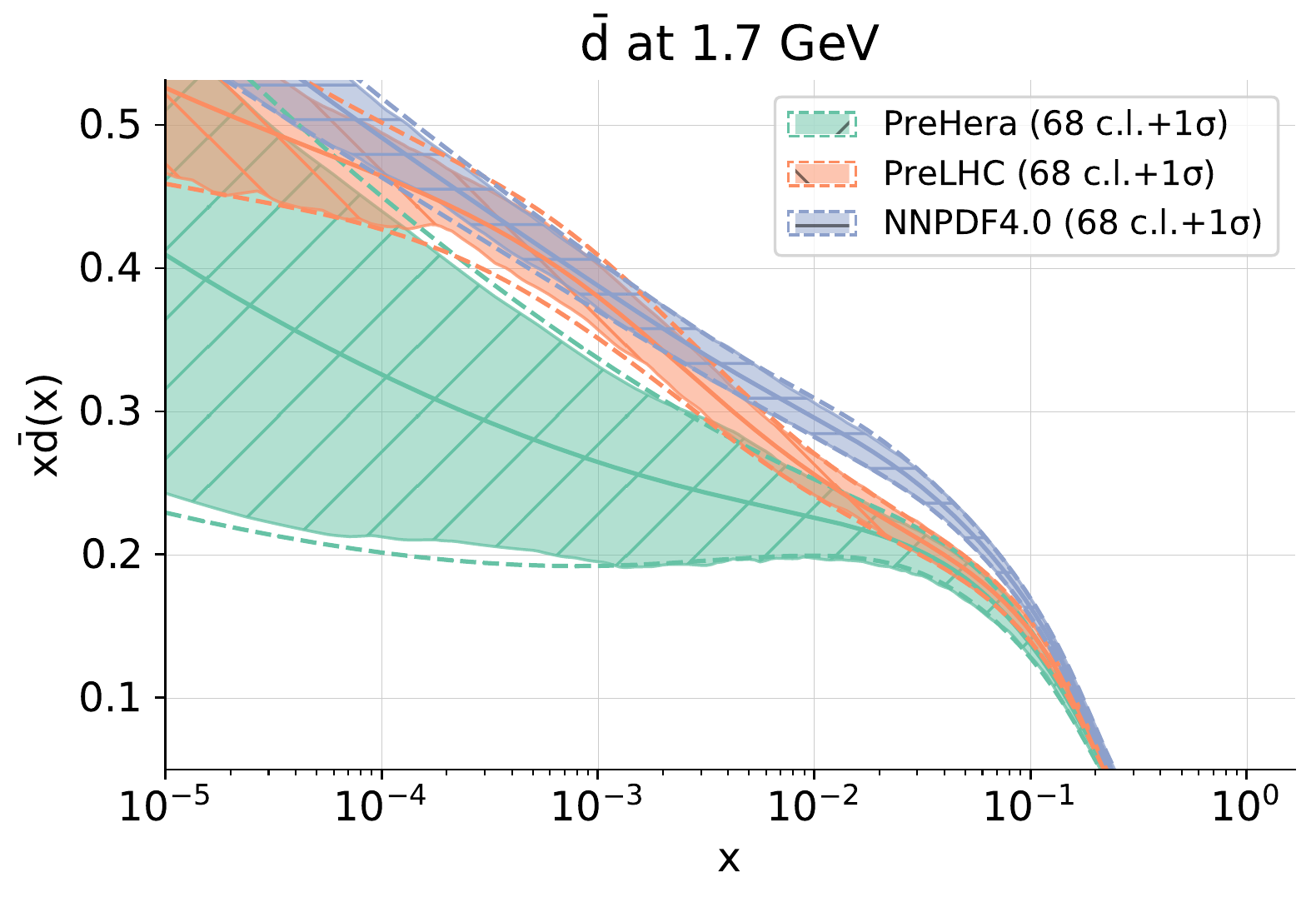}
  \includegraphics[width=0.49\textwidth]{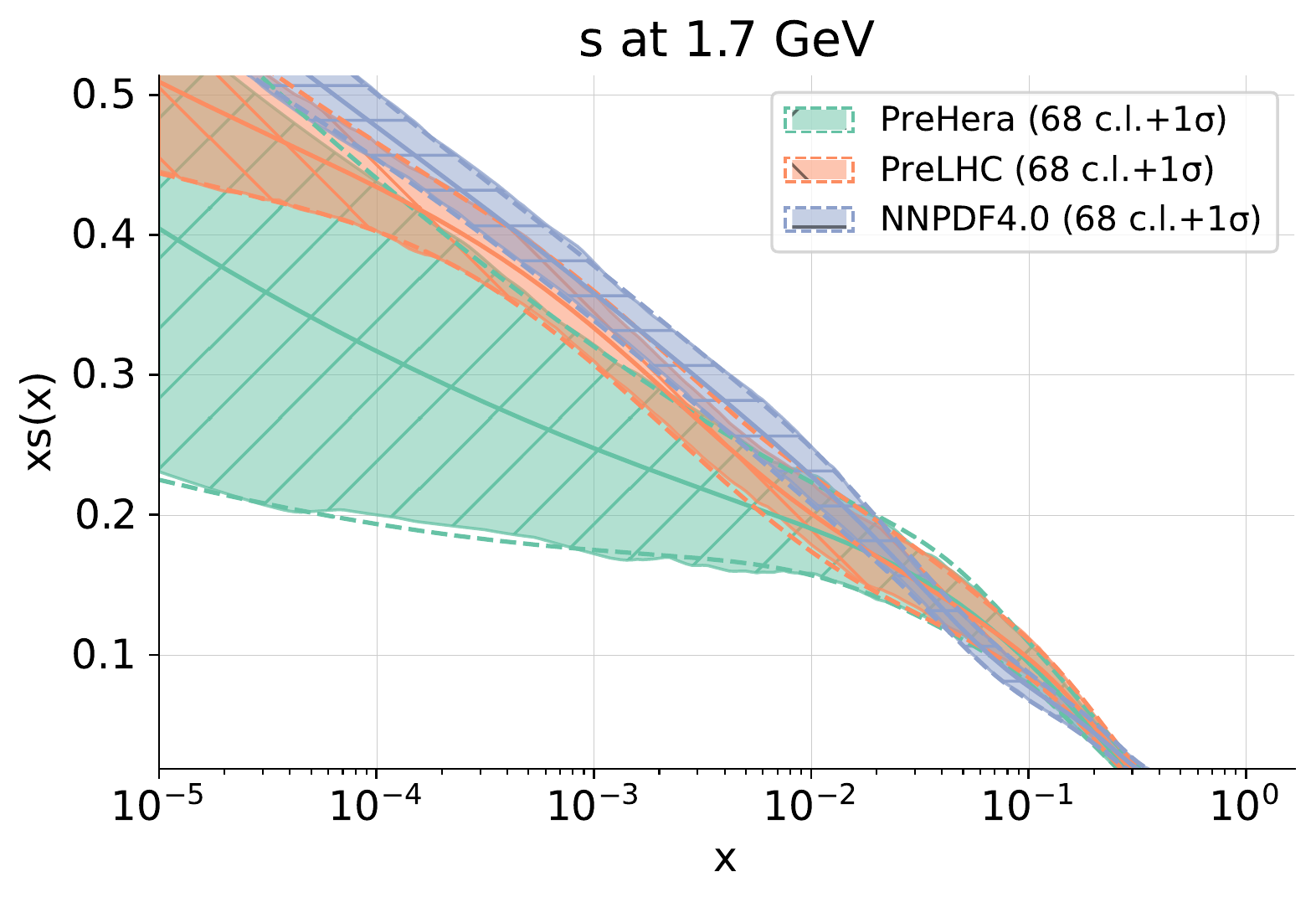}
  \includegraphics[width=0.49\textwidth]{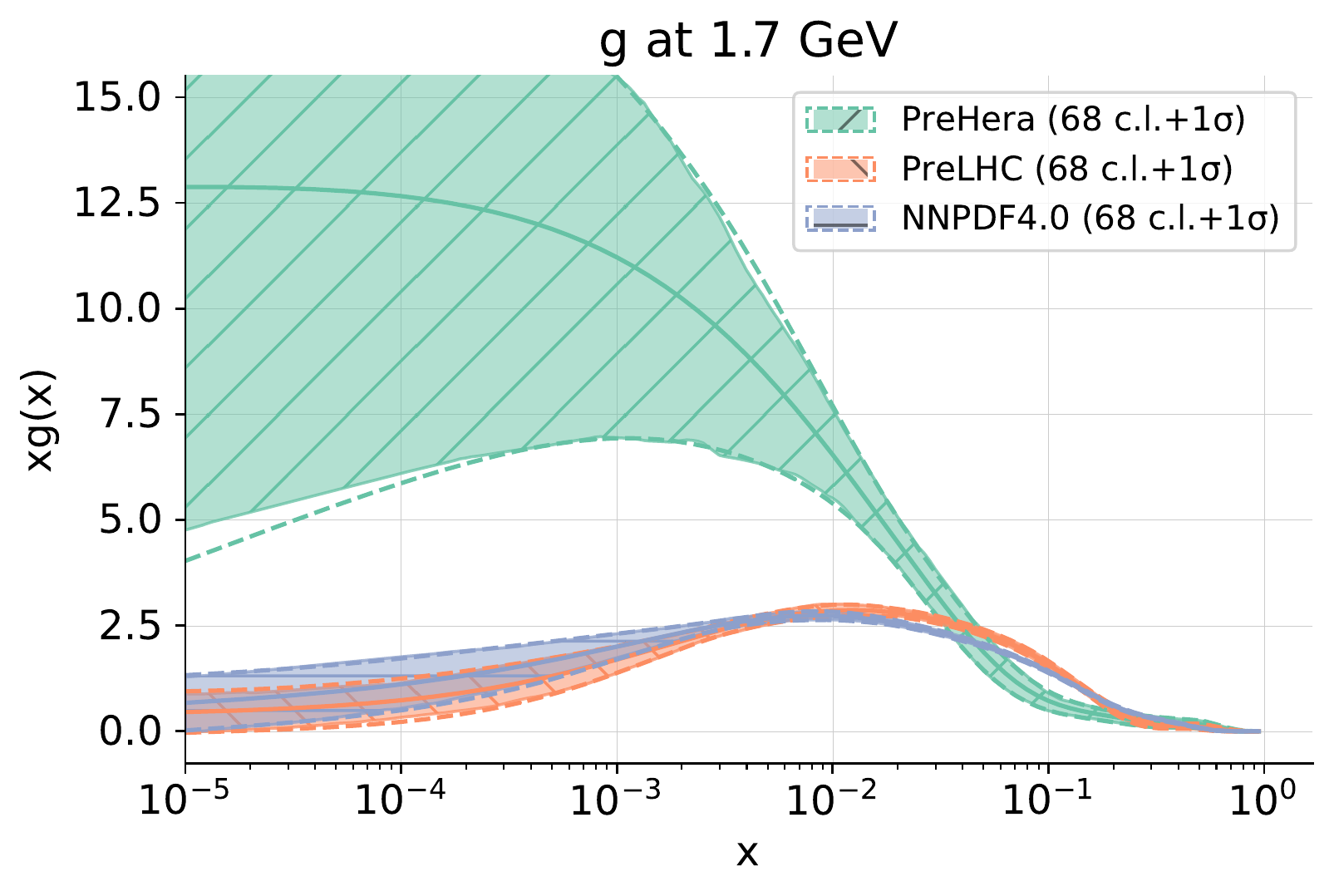}
  \caption{\small Comparison of the pre-HERA, pre-LHC and NNPDF4.0
    PDFs. The up (top left), antidown (top right), strange (bottom
    left) and gluon (bottom right) PDFs are shown at the parametrization
    scale $Q=1.7$~GeV.
   }
    \label{fig:unpolpdfs}
\end{center}
\end{figure}

The relevant information from Table~\ref{tab:chi2unp}
is contained in the $\chi^2$ values corresponding to predictions,
i.e. computed from data not included in the determination of the
corresponding PDF sets. These are shown in italic (without PDF
uncertainties) and boldface (with PDF uncertainties). These values
show that the future
test is impressively successful, and the criterion
Eq.~(\ref{eq:criterion}) is clearly satisfied.
Before inclusion of PDF uncertainties
(values in italic),
the $\chi^2$ of future test PDFs is of order 20, but after their
inclusion (values in boldface)
it becomes of order one, and in fact almost identical to
that of fitted PDFs (without uncertainties included). This means that
PDF uncertainties increase on average by more than two orders of
magnitude in the extrapolation region, and that this increase is
exactly of the size required in order to account for the missing information
contained in the excluded data. This is especially remarkable for
  the $\chi^2$ of the pre-HERA PDFs when compared to the post-HERA
  dataset, whose value is hardly larger than that found in the
  NNPDF4.0 fit itself. In fact, inspection of $\chi^2$ values for
  individual experiments~\cite{NNPDF40} shows that the $\chi^2$
  values after inclusion of PDF uncertainties are all very close to
  the values found in the  NNPDF4.0 fit (without PDF uncertainties),
  while, especially for the pre-HERA PDFs, some of the values found
  withiut PDF uncertainties,
  especially for experiments in far extrapolation regions, can be up
  to two orders of magnitude higher, as we will discuss below for some
  specific examples.  
  This means that uncertainties are quite
  reliable even in far extrapolation regions.

Note that the $\chi^2$ values for the
data included in the fit (which, as mentioned, do not have a
statistical meaning strictly) are
almost unchanged when PDF uncertainties are included. This is due to
the fact that, for data included in the fit,
   the
  PDF uncertainty is generally rather smaller than that of data used for its
  determination, because PDFs combine the information coming from many
  datapoints. Hence for these data the experimental uncertainty is
  dominant, and this explains why $\chi^2$ values are almost
  unaffected by the inclusion of PDF uncertainties.

\begin{figure}[t]
\begin{center}
  \includegraphics[width=0.49\textwidth]{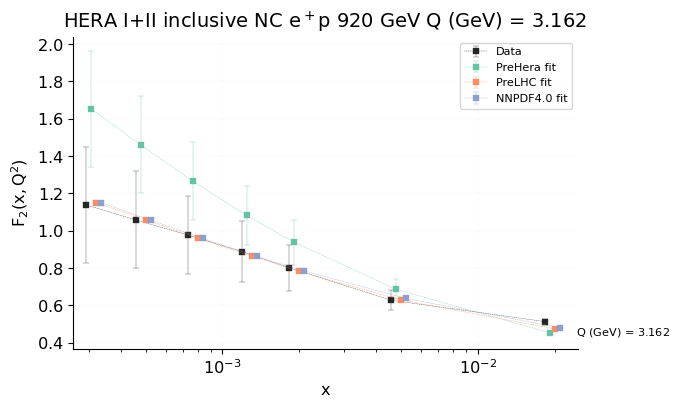}
  \includegraphics[width=0.49\textwidth]{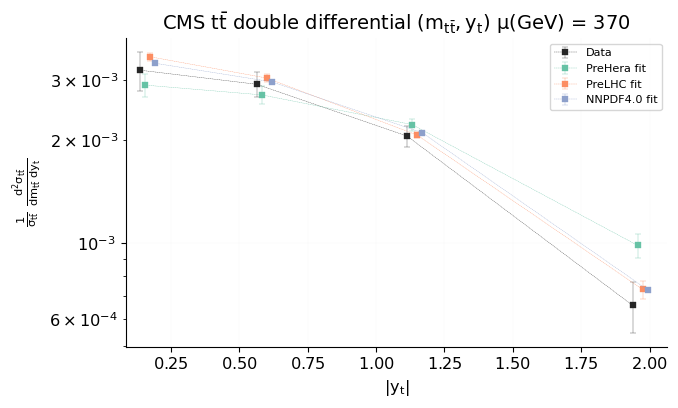}
  \includegraphics[width=0.49\textwidth]{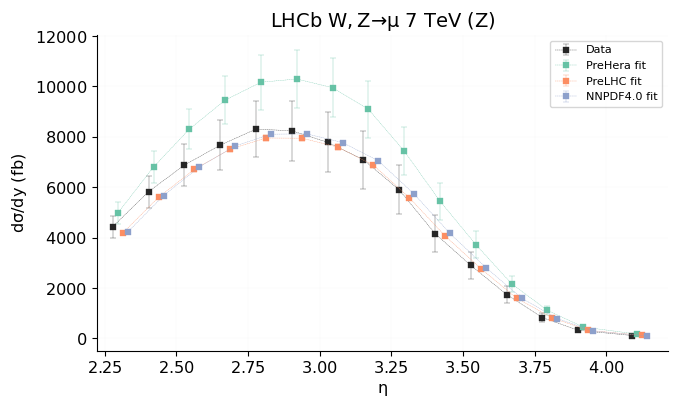}
  \includegraphics[width=0.49\textwidth]{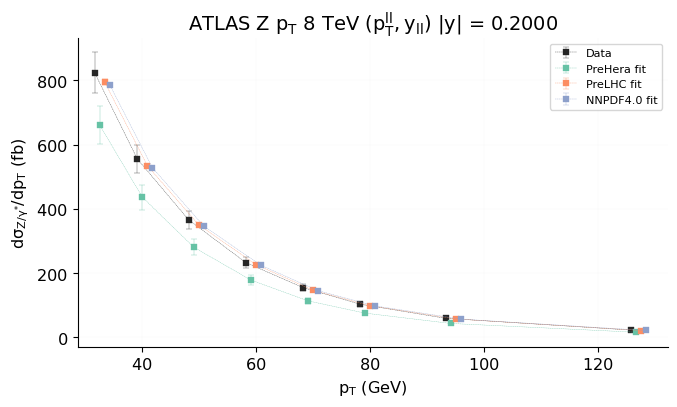}
  \caption{\small Comparison to experimental data of predictions
    obtained using the pre-HERA, pre-LHC and NNPDF4.0
    PDF sets.  The diagonal PDF uncertainties and data uncertainties
    are shown. Illustrative results are displayed for the HERA $F_2$ proton
    structure function~\cite{Abramowicz:2015mha} (top left) the CMS double
    differential top pair distribution~\cite{Sirunyan:2017azo} (top right),
    the LHCb  $Z$ rapidity distribution~\cite{Aaij:2015gna}
    (bottom left) and the ATLAS $Z$ $p_T$ distribution~\cite{Aad:2015auj}
    (bottom right).
   }
    \label{fig:unpoldata}
\end{center}
\end{figure}
  A more detailed picture can be obtained by looking at individual
  PDFs. Specifically, in Fig.~\ref{fig:unpolpdfs} the up, antidown,
  strange quark and gluon PDFs found in the  two future test fits are
  compared to NNPDF4.0. It is clear that PDFs are generally compatible
  within errors, with differences exceeding the one sigma level
  in a limited set of cases (as expected given that one sigma
  corresponds to a $68\%$ confidence level). This is true both for
  PDFs determined relatively accurately (such as the up quark in the
  pre-LHC fit), or those affected by large uncertainties (such as the
  gluon in the pre-HERA fit).

  The case of the pre-HERA gluon is
  especially remarkable: despite the very large uncertainty, the
  future test correctly extrapolates the small $x$ rising trend. In
  fact, with hindsight this trend can be seen in some pre-HERA DIS data (especially
  from NMC), though historically it was overlooked. Of course, the
  methodology that we are future testing has been hyperoptimized on a
  dataset that does contain small-$x$ data, so the pre-HERA future
  test cannot be taken as
  evidence that the rise of the structure function could have been
  predicted without the HERA data. However, it does suggest that an
  unbiased inspection of this data should have at least have suggested
  this possibility. 

  In Fig.~\ref{fig:unpoldata} we also show the comparison of
  predictions obtained using either NNPDF4.0, or the two future test
  PDF sets, to data for some physical observables: the HERA $F_2$
  structure function~\cite{Abramowicz:2015mha}, the CMS double-differential
  distribution of top pairs~\cite{Sirunyan:2017azo}, the LHCb $W/Z$ rapidity
  distribution~\cite{Aaij:2015gna} and the ATLAS $Z$ $p_T$
  distribution~\cite{Aad:2015auj}. These observables are sensitive to a wide
  array of PDF features which are poorly constrained by pre-HERA
  data. Specifically, the HERA structure function probes the small $x$ gluon,
  the top pair and $Z$ $p_T$ distributions depend on the large $x$
  gluon and the LHCb large-rapidity $W$ and $Z$ cross sections probe
  the flavor separation at  large $x$. In each case we clearly see hat
  the PDF uncertainty on the prediction using future test PDF
  correctly accounts for the deviation from the data, despite the very
  substantial amount of extrapolation which is required in all cases
  shown. This is manifestly seen when inspecting the corrresponding
  $\chi^2$ values. For instance, the 
  $\chi^2$ values per datapoint for the pre-HERA fit are, for the HERA
  I+II inclusive NC $e^+p$ 920 GeV  data, $\chi^2=44.2$
  without PDF uncertainties and ${\chi^2}^{\rm PDF}=1.29$ when PDF uncertainties
  are included (fitted NNPDF4.0 value $\chi^2=1.31$); for CMS $t\bar{t}$ 2D $2\ell$ $(m_{t\bar{t}},y_{t})$ 
  8~TeV, they are respectively
  $\chi^2=5.125$ and
  ${\chi^2}^{\rm PDF}=1.15$ (fitted NNPDF4.0 value $\chi^2=0.89$); for LHCb $W,Z \to \mu$ 7 TeV, $\chi^2=15.37$ and
  ${\chi^2}^{\rm PDF}=1.71$ (fitted NNPDF4.0 value $\chi^2=1.96$); and for ATLAS $Z$ $p_T$ 8 TeV $(p_T^{ll},y_{ll})$,
  $\chi^2=28.69$  and
  ${\chi^2}^{\rm PDF}=0.997$ (fitted NNPDF4.0 value $\chi^2=0.89$).  In
  all cases, the inclusion of the PDF uncertainties brings down the
  $\chi^2$ to a value that is close to that found when the data are
  fitted, and the criterion Eq.~(\ref{eq:criterion}) is very well
  satisfied, as the figures show graphically.


\begin{figure}[t]
\begin{center}
  \includegraphics[width=0.49\textwidth]{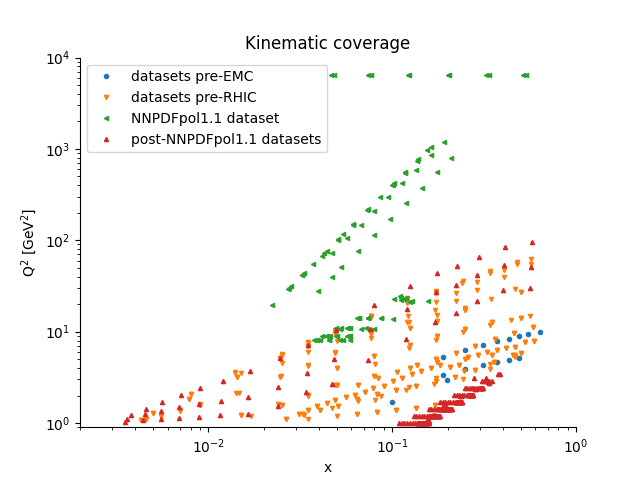}
  \includegraphics[width=0.49\textwidth]{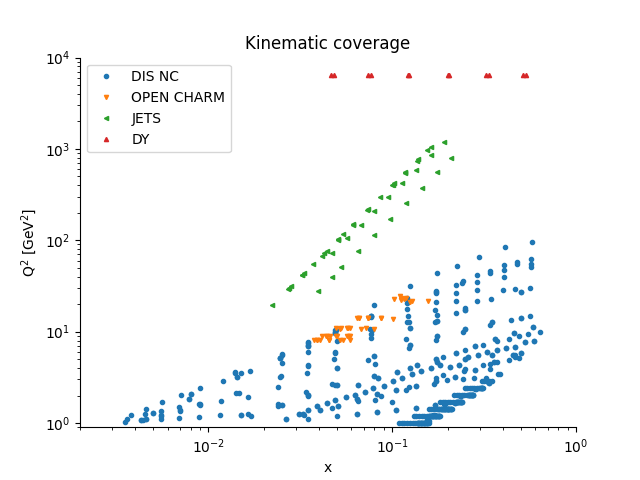}
  \caption{\small Same as Fig.~\ref{fig:scatter} but now for polarized
    PDFs.
   }
    \label{fig:scatterpol}
\end{center}
\end{figure}
\section{Polarized PDFs and the ``spin crisis''}
\label{sec:CTpol}

We now turn to longitudinally polarized parton distributions
(polarized PDFs, henceforth), defined as
\begin{equation}
  \label{eq:qpol}
\Delta f_i(x,Q^2)=
f^{\uparrow\uparrow}_i(x,Q^2)-f^{\uparrow\downarrow}_i(x,Q^2),
\end{equation}
where $f^{\uparrow\uparrow}_i$ ($f^{\uparrow\downarrow}$) is the 
$i$-th species of PDF  when the parton's spin is parallel
(antiparallel) to that of its parent hadron, so that  unpolarized PDFs are 
$f_i(x,Q^2)=
f^{\uparrow\uparrow}_i(x,Q^2)+f^{\uparrow\downarrow}_i(x,Q^2)$.

Polarized PDFs are based on rather more scarce experimental
information than their unpolarized counterparts. Not only a much
smaller number of data and processes is available, but also these are
typically affected by significantly larger uncertainties. Furthermore
fully correlated systematics are often not available, so statistical and
systematic uncertainties must be added in quadrature, thereby leading
to even larger uncertainties. Extrapolation problems are accordingly
more serious for polarized PDFs.

\begin{table}[!t]
\renewcommand*{\arraystretch}{1.60}
\scriptsize
\centering
\begin{tabularx}{.49\textwidth}{|Xcr|}
\toprule
{\bf pre-EMC}
    & Ref. 
    & $N_{\rm{dat}}$ \\
\midrule
SLAC (1978) $A_1^p$
    & \cite{Alguard:1978gf}
    &   4 \\
SLAC (1983) $A_1^p$
    & \cite{Baum:1983ha}
    &  14 \\
\midrule
Total pre-EMC
    & 
    &  18 \\
\midrule
{\bf pre-RHIC}
    & Ref.    
    & $N_{\rm{dat}}$ \\
\midrule
EMC $A_1^p$
    & \cite{Ashman:1989ig}
    &  10 \\
SMC $A_1^p$
    & \cite{Adeva:1998vv}
    &  12 \\
SMC $A_1^d$
    & \cite{Adeva:1998vv}    
    &  12 \\
SMC (low-$x$) $A_1^p$
    & \cite{Adeva:1999pa}
    &   8 \\
SMC (low-$x$) $A_1^d$
    & \cite{Adeva:1999pa}
    &   8 \\
E143 $A_1^p$
    & \cite{Abe:1998wq}
    &  25 \\
E143 $A_1^d$
    & \cite{Abe:1998wq}
    &  25 \\
E154 $A_1^n$
    & \cite{Abe:1997cx}
    &  11 \\
E155 $g_1^p$
    & \cite{Anthony:2000fn}
    &  20 \\
E155 $g_1^d$
    & \cite{Anthony:2000fn}
    &  20 \\
COMPASS (2007) $A_1^d$
    & \cite{Alexakhin:2006oza}
    &  15 \\
COMPASS (2010) $A_1^p$
    & \cite{Alekseev:2010hc}
    &  15 \\
HERMES  (1997) $A_1^n$
    & \cite{Ackerstaff:1997ws}
    &   8 \\
HERMES $A_1^p$
    & \cite{Airapetian:2007mh}
    &  28 \\
HERMES $A_1^d$
    & \cite{Airapetian:2007mh}
    &  28 \\
\midrule
Total pre-RHIC
    & 
    & 245 \\
\bottomrule
\end{tabularx}
\hfill
\renewcommand*{\arraystretch}{1.65}
\begin{tabularx}{.49\textwidth}{|Xcr|}
\toprule
{\bf NNPDFpol1.1}
& Ref.
& $N_{\rm dat}$ \\
\midrule
COMPASS open charm $A_{LL}$
    & \cite{Adolph:2012ca}
    &  45 \\
STAR 1-jet (2005)  $A_{LL}$
    & \cite{Adamczyk:2012qj}
    &  10 \\
STAR 1-jet (2006)  $A_{LL}$
    & \cite{Adamczyk:2012qj}
    &   9 \\
STAR 1-jet (2009)  $A_{LL}$
    & \cite{Adamczyk:2014ozi}
    &  22 \\
PHENIX 1-jet       $A_{LL}$
    & \cite{Adare:2010cc}
    &   6 \\
\midrule
Total NNPDFpol1.1
    &
    &  92 \\
\midrule
{\bf post-NNPDFpol1.1}
    & Ref.
    & $N_{\rm{dat}}$ \\
\midrule
COMPASS (2015) $A_1^p$
    & \cite{Adolph:2015saz}
    &  51 \\
COMPASS (2016) $A_1^d$
    & \cite{Adolph:2016myg}
    &  43 \\
JLAB-EG1-DVCS $g_1^p$
    & \cite{Prok:2014ltt}
    &  9  \\
JLAB-EG1-DVCS $g_1^d$
    & \cite{Prok:2014ltt}
    &  9  \\
JLAB-E93-009 $A_1^p$
    & \cite{Guler:2015hsw}
    &  62 \\
JLAB-E93-009 $A_1^d$
    & \cite{Guler:2015hsw}
    &  86 \\
JLAB-E06-014 $A_1^n$
    & \cite{Parno:2014xzb}
    &   2 \\
STAR 2-jet (2009)  $A_{LL}$
    & \cite{Adamczyk:2016okk}
    &  14 \\
STAR 1-jet (2012)  $A_{LL}$
    & \cite{Adam:2019aml}
    &  14 \\
STAR 2-jet (2012)  $A_{LL}$
    & \cite{Adam:2019aml}
    &  42 \\
\midrule
Total postNNPDFpol1.1
    &
    & 332 \\
\midrule
Grand Total
    &
    & 687 \\
\bottomrule
\end{tabularx}

\vspace{0.3cm}
\caption{\small
  The pre-EMC, pre-RHIC, NNPDFpol1.0 and post-NNPDFpol1.1 datasets: for each
  experiment the reference to the original publication and
  the number of datapoints is given. Pre-EMC polarized PDFs are
  fitted to pre-NMC, pre-RHIC PDFS are fitted to the pre-EMC
  and pre-RHIC data, and NNPDFpol1.1 PDFs are fitted to the union of these
  three datasets. Post-NNPDFpol1.1 PDFs are only used for future
  testing. }
\label{tab:poldat}
\end{table}

\subsection{Polarized PDFs and data}
\label{sec:poldat}

Until quite recently, polarized PDFs where determined purely using
neutral-current deep-inelastic scattering on proton,
deuteron and
neutron (i.e., Helium) targets, from the
polarized structure function $g_1$ directly, or the polarized
asymmetries from which the structure function is extracted 
(see~\cite{Aidala:2012mv} for a review). More recently these were supplemented by open charm and semi-inclusive hadron production and,
more importantly, by precious little data on $W$, inclusive jet and pion
production from RHIC, see~\cite{Ethier:2020way} and references therein. The dataset for the most recent determination of polarized
PDFs based on the NNPDF methodology, NNPDFpol1.1~\cite{Nocera:2014gqa}, is
shown in Fig.~\ref{fig:scatterpol}. We have performed a three-fold
future test of this PDF set and methodology, by comparing it to
pre-EMC and pre-RHIC datasets, and also to a post-NNPDF1.1 dataset. This dataset
includes some more recent data, which appeared after  the
original NNPDFpol1.1 PDF determination. Some of these data were discussed,
in the context of NNPDF fits, in
Refs.~\cite{Nocera:2015vva,Nocera:2017wep} without leading to a fully updated PDF
release. The datasets are all shown in Fig.~\ref{fig:scatterpol}, and
also listed in Table~\ref{tab:poldat}. Because the NNPDFpol1.1
methodology is essentially the same as the NNPDF3.1 methodology, the
future tests presented here also effectively  test the NNPDF3.1 methodology.

\begin{figure}[t]
\begin{center}
  \includegraphics[width=0.49\textwidth]{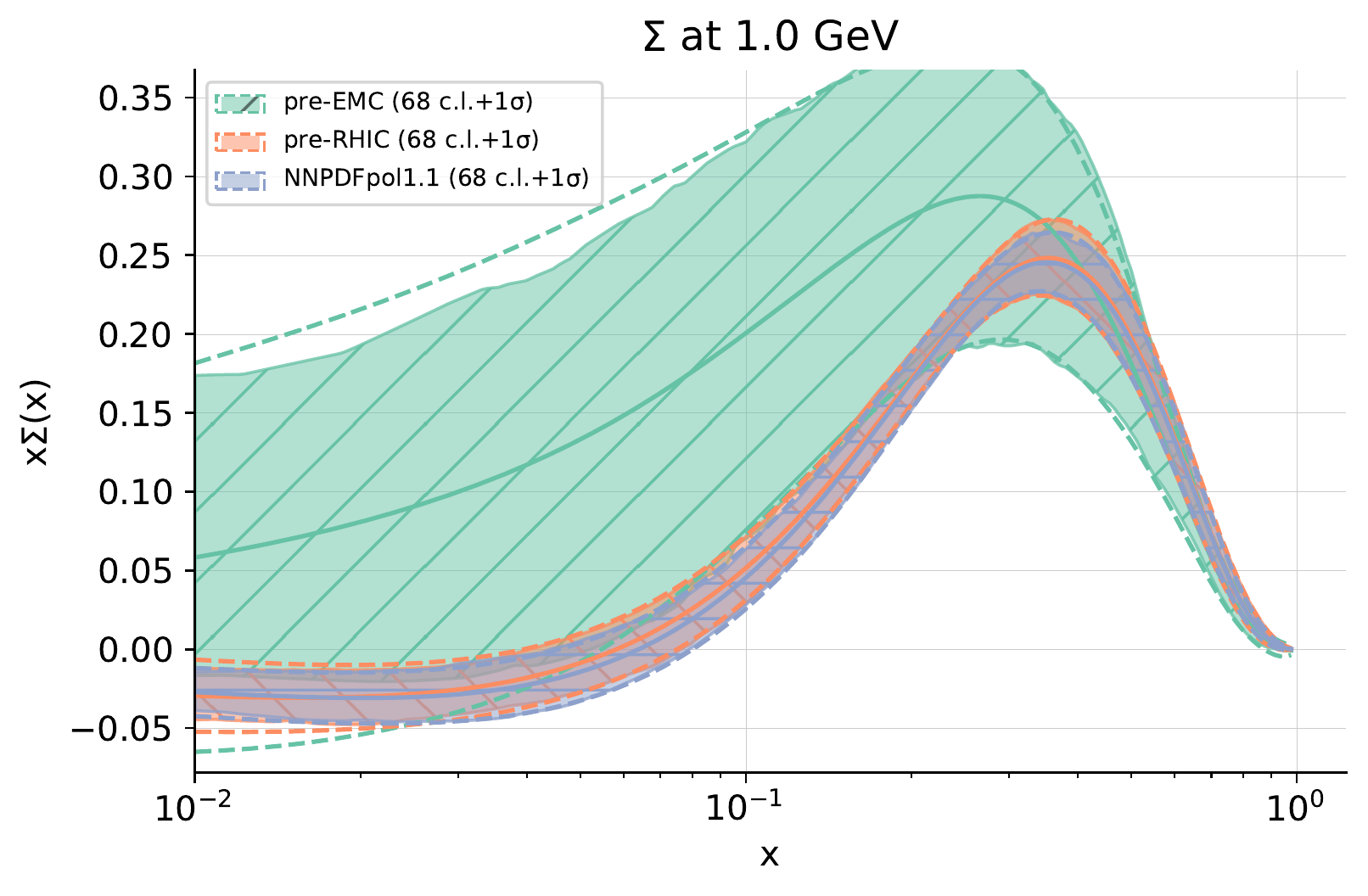}
  \includegraphics[width=0.49\textwidth]{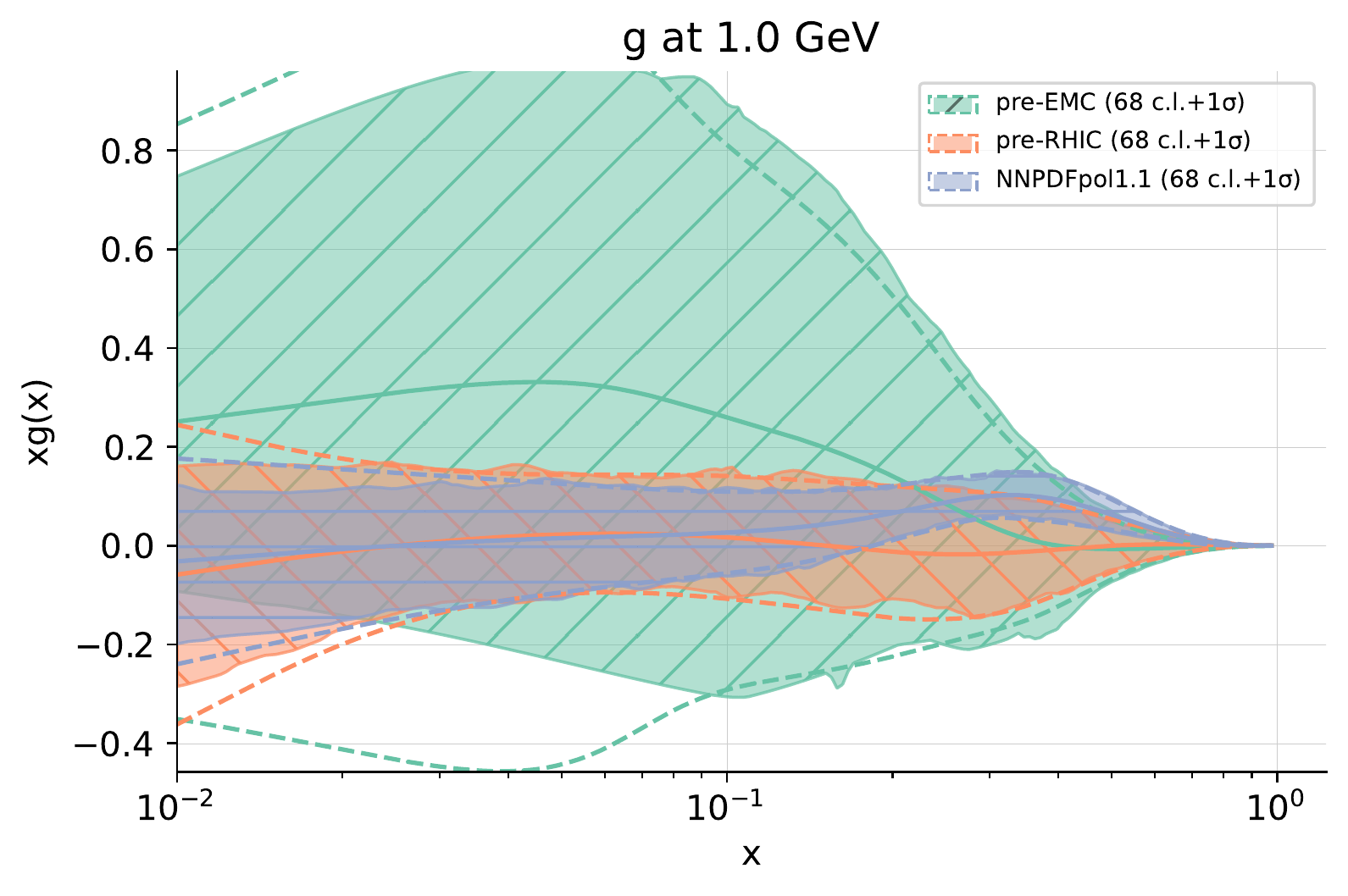}\\
  \includegraphics[width=0.49\textwidth]{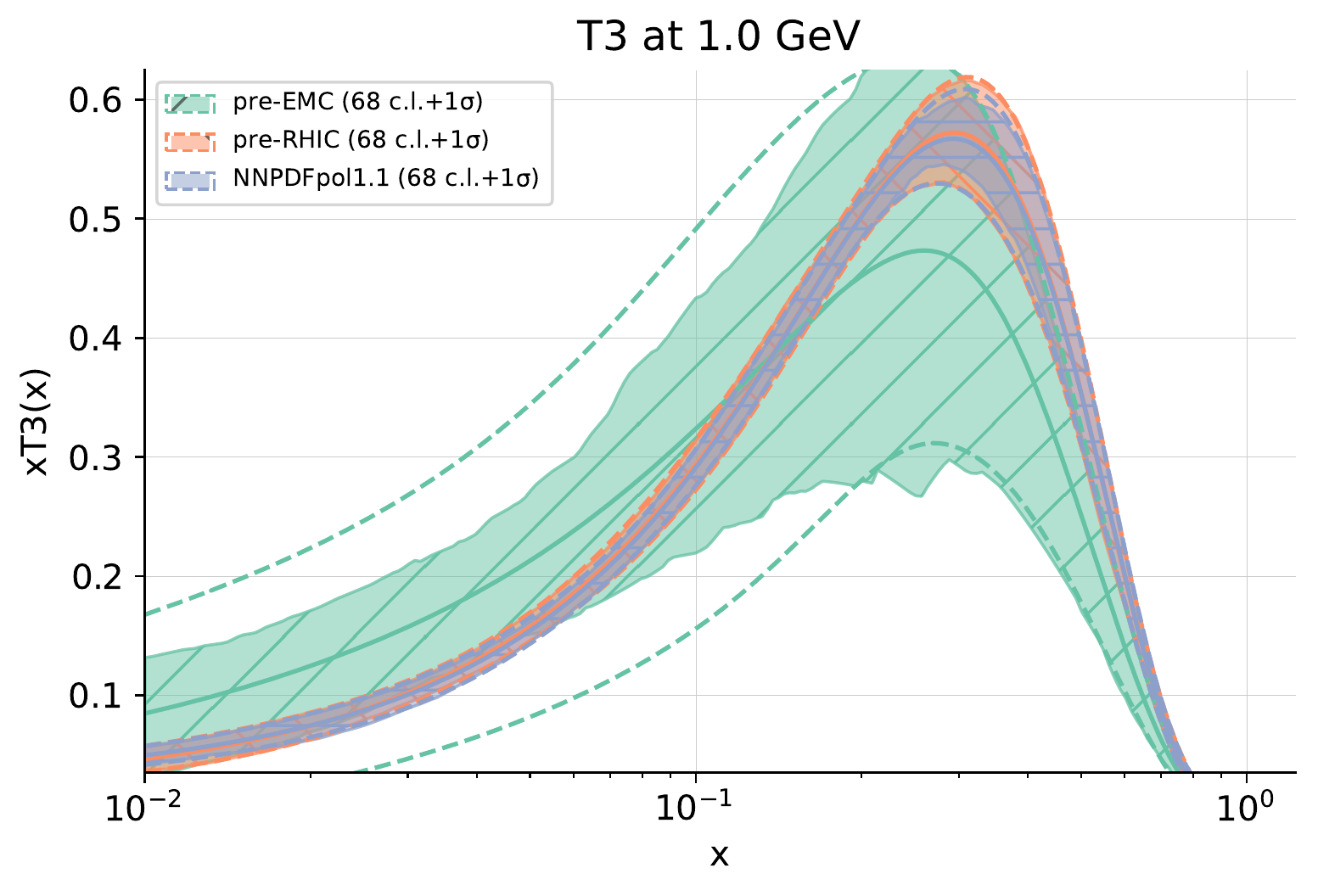}
  \includegraphics[width=0.49\textwidth]{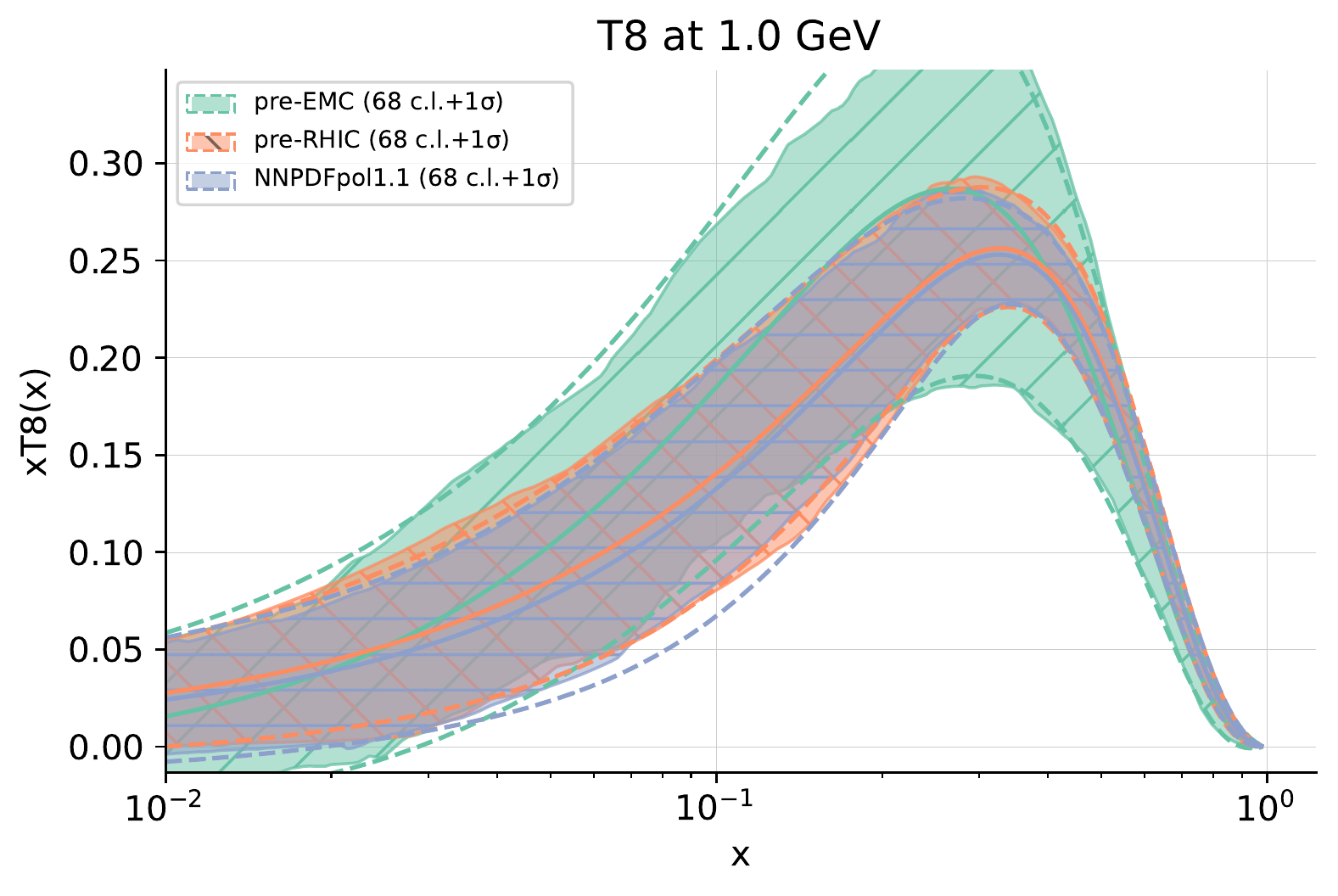}
  \caption{\small Comparison of the pre-EMC, pre-RHIC and NNPDFpol1.1
    PDFs. The singlet (top left), gluon (top right), triplet  (bottom
    left) and octet (bottom right) PDF combinations (see text)
    are shown at the parametrization
    scale $Q=1$~GeV.
   }
    \label{fig:polpdfs}
\end{center}
\end{figure}
A PDF determination based purely on neutral-current DIS cannot
disentangle quark and antiquark PDFs, and only allows for a
determination of their sum:
\begin{equation}
  \label{eq:cev}
  \Delta q_i^+(x,Q^2)= \Delta q_i(x,Q^2)+\bar q_i(x,Q^2)
  \end{equation}
for each flavor $i$.
Furthermore, DIS on proton targets only
determines a fixed linear combination of quark flavors, i.e., in
practice a fixed linear combination of up, down and strange
contributions since the contribution of heavy quarks is negligible, at
least within present-day accuracy. If also deuteron or neutron targets
are available, it is then possible to determine separately the triplet
combination
\begin{equation}
  \label{eq:trip}
  \Delta T_3(x,Q^2)= \Delta u^+(x,Q^2)-\Delta d^+(x,Q^2)\, .
\end{equation}
However, it is not a priori possible to disentangle the two linear
combinations orthogonal to it, namely the singlet
\begin{equation}
  \label{eq:sing}
  \Delta \Sigma(x,Q^2)= \Delta u^+(x,Q^2)+\Delta d^+(x,Q^2)+\Delta s^+(x,Q^2)
\end{equation}
and octet
\begin{equation}
  \label{eq:oct}
  \Delta T_8(x,Q^2)= \Delta u^+(x,Q^2)+\Delta d^+(x,Q^2)- 2 \Delta s^+(x,Q^2)\, .
\end{equation}
However, the first moments (i.e. integrals) of triplet and
octet can be extracted from weak decays. If only DIS on proton targets
is used, this is then the only information available on flavor
separation. 

The pre-EMC dataset, which only includes proton DIS data, only allows for a
determination of quark and gluon (the latter from the $Q^2$ dependence),
with minimal information on flavor separation from first moments.
The pre-RHIC dataset, including proton, deuteron and neutron DIS data
allows for a more detailed flavor separation, and a reasonable
determination of the triplet. The NNPDFpol1.1 allows for a
determination of individual flavor and antiflavors (from $W$
production) with jet production providing a direct handle on the
gluon. These datasets are therefore hierarchical in terms of the flavor
separation they allow:  pre-EMC and pre-RHIC correspond respectively
to far and near extrapolation in flavor space.

\begin{figure}[t]
\begin{center}
  \includegraphics[width=0.49\textwidth]{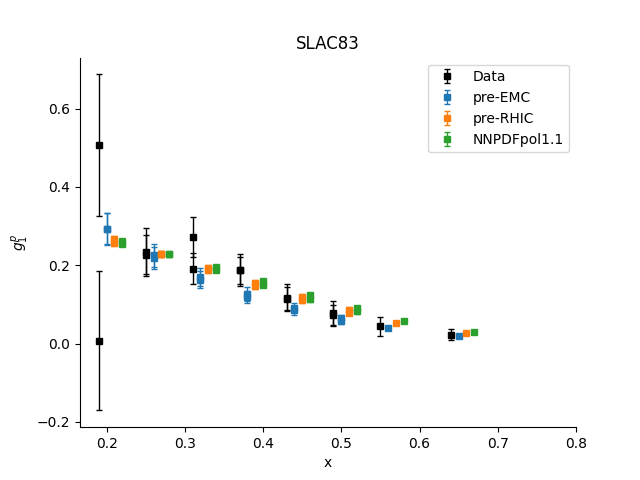}
  \includegraphics[width=0.49\textwidth]{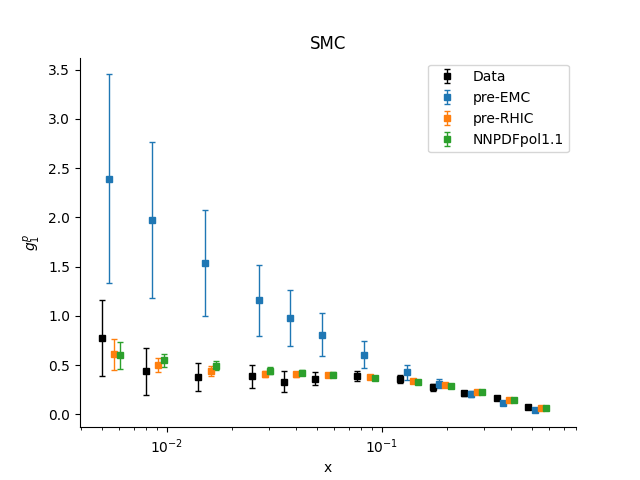}\\
  \includegraphics[width=0.49\textwidth]{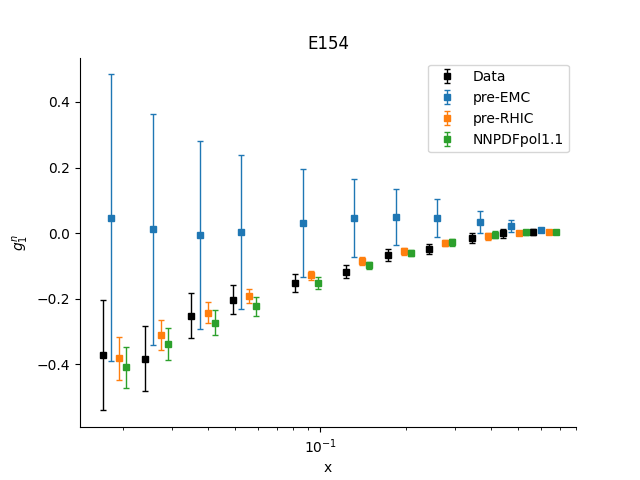}
  \includegraphics[width=0.49\textwidth]{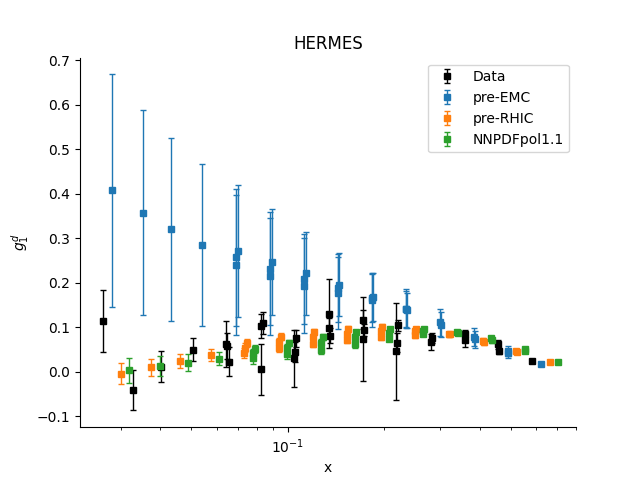}
  \caption{\small Comparison to experimental data of predictions
    obtained using the pre-EMC, pre-RHIC and NNPDFpol1.1
    PDF sets.  The diagonal PDF uncertainties and data uncertainties
    are shown. Illustrative results are displayed for
    the proton structure function $g_1^p$ measured at SLAC pre-EMC (top
    left)~\cite{Baum:1983ha} and measured by SMC (top right)~\cite{Adeva:1998vv}, as well as for the E154
    measurement of the neutron structure function $g_1^n$  (bottom left)~\cite{Abe:1997cx} and the
    HERMES measurement of the deuterium structure function  $g_1^d$
    (bottom right)~\cite{Airapetian:2007mh}. 
   }
    \label{fig:poldata}
\end{center}
\end{figure}
As is apparent from Fig.~\ref{fig:scatterpol}, these datasets are
also hierarchical in terms of coverage in $(x,Q^2)$ space. More
coverage in $Q^2$ allows for a better extraction of the gluon, which
is determined mainly by the scale dependence, while more coverage in $x$
allows for a better determination of the first moment of individual
PDFs, and in particular the first moment of the singlet combination.
Unlike the triplet and octet, the singlet is not determined by weak decays,
and has a simple physical interpretation as the fraction of the parent
hadron's spin carried by quarks (up to field theoretical complications
related to the axial anomaly~\cite{Altarelli:1988nr}). So also in
terms of $x$-coverage, gluon determination, and determination of the
first moments, the pre-EMC and pre-RHIC datasets correspond to far and
near extrapolation. 

In fact, this
is the reason why specifically a pre-EMC dataset was chosen for future
testing. Indeed, when the
EMC data~\cite{Ashman:1989ig} were originally
published, a first determination of the singlet first
moment was possible. The result turned out to be surprising:
the first moment  was found to be  rather smaller than
expected, and in fact compatible with zero within uncertainties
--- which would mean (at least naively)
that the proton spin is not carried by quarks. This has been often
referred to as the ``proton spin crisis''~\cite{Leader:1988vd}. Hence, just like in the
case of the rise of the structure function at HERA, it is natural to
wonder what a contemporary polarized methodology would make of pre-EMC
data.

\begin{figure}[t]
\begin{center}
  \includegraphics[width=0.49\textwidth]{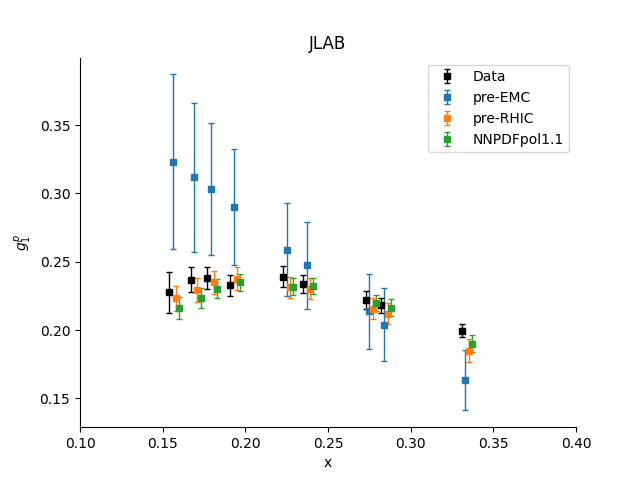}
  \includegraphics[width=0.49\textwidth]{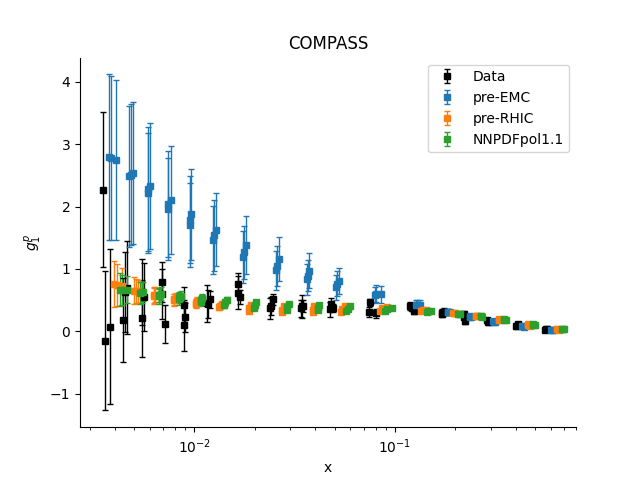}
  \caption{\small Same as Fig.~\ref{fig:poldata}, but now comparing to
    post-NNPDFpol1.1 data. Results are shown for
    $g_1^p$ measured at JLAB (left)~\cite{Prok:2014ltt} and COMPASS (right)~\cite{Adolph:2015saz}. 
   }
    \label{fig:poldatab}
\end{center}
\end{figure}
Finally, we have also performed a future test with post-NNPDFpol1.1
data because these significantly extend the precision and also somewhat
enlarge the kinematic range of the NNPDFpol1.1 dataset: in a sense
these are the first polarized data that provide a first step towards
the direction of precision polarized PDFs. However, we have
future-tested them using the NNPDFpol1.1 methodology because this is the
most recent polarized methodology that we have. In this sense, this
future test confirms (or disproves) that the NNPDFpol1.1 methodology is
still adequate, even in view of this more recent data.

\begin{table}[t]
\renewcommand*{\arraystretch}{1.60}
\scriptsize
\centering
\begin{tabularx}{\textwidth}{|X|c|ccc||ccc|}
\toprule
                           & $n_{\rm dat}$ & NNPDFpol1.1 & pre-RHIC & pre-EMC & NNPDFpol1.1 & pre-RHIC & pre-EMC \\
\midrule
pre-EMC dataset            &    18       & 0.53        &  0.53    & 1.09    & (0.53)      & (0.52)   & (0.77)  \\
pre-RHIC dataset           &   245       & 0.75        &  0.77    &
{\it 20.4}    & (0.64)      & (0.67)   &  {\bf 0.51}   \\
NNPDFpol1.1 dataset        &    92       & 1.37        &  {\it 1.66}    &
{\it 7.36}    & (1.35)      &  {\bf 1.38}    &  {\bf 1.40}   \\
post-NNPDFpol1.1 dataset   &   332       & {\it 2.58}        &  {\it 2.54}    &
{\it 8.46}    &  {\bf 0.97}       &  {\bf 0.96}    &  {\bf 0.90}   \\
\midrule
Total dataset              &   687       & 1.54        &  1.58    & 13.2    & (0.88)      & (0.89)   & (0.81)  \\
\bottomrule
\end{tabularx}

\vspace{0.3cm}
\caption{\small The $\chi^2$ per datapoint for the NNPDFpol1.0 PDF determination
and its pre-EMC and pre-RHIC future tests. Values are shown for the
four datasets displayed in Fig.~\ref{fig:scatterpol}. In the left table
all values are computed using only the experimental covariance matrix,
while in the right table all values are computed by also including PDF
uncertainties. Values with PDF uncertainties included computed for data
included in the fit, shown in parenthesis, do not have a strict
statistical meaning (see text). All numbers in italic (without PDF
uncertainty included) and in boldface (with PDF uncertainty included)
are predictions.}
\label{tab:chi2pol}
\end{table}
\label{sec:polpdf}
\subsection{Future testing polarized PDFs}
\label{sec:emceic}
The polarized future test tests the NNPDFpol1.1
methodology~\cite{Nocera:2014gqa}. This is quite similar to the
methodology used for unpolarized NNPDF sets until the most recent published
  NNPDF3.1~\cite{Ball:2017nwa}. It differs from the methodology whose
  future test was discussed in Section~\ref{sec:40}, essentially because
  it is not obtained from an automatic hyperoptimization procedure, so
methodological details, such as the choice of neural network
architecture, are based on previous experience rather then being
optimized automatically. A peculiar aspect of the polarized fitting
methodology is the need to assume an underlying set of unpolarized
PDFs, since experimental data are usually presented for polarized
asymmetries. Polarized PDF determinations presented in this section
assume the published NNPDF3.1 as an underlying unpolarized set.

As in the unpolarized case, we have computed $\chi^2$ values for all
datasets and PDF sets, both with and without PDF uncertainties
included. Results are collected in Table~\ref{tab:chi2pol}. An
important caveat in the polarized case is that correlated systematic
uncertainties are mostly not available, except for the most recent
experiments. Hence,  statistical and systematic
uncertainties have to be added in quadrature, and the covariance
matrix must be taken to be diagonal. We will come back to this point
shortly. PDF uncertainties, on the other hand, are always included with
all correlations fully accounted for.

As in the
unpolarized case, the relevant comparison is between $\chi^2$ values
for data not included in the fit, before (in italic) and after (in
boldface) the inclusion of PDF uncertainties. As in the unpolarized
case, we find that
all $\chi^2$ values with PDF uncertainties included
are of order one, thus showing that the future test is successful. The
pre-EMC case is especially remarkable, with a $\chi^2$ value that, for
data not used for fitting, is of order
thirteen per datapoint, and decreases to a value of order one once the
PDF uncertainty is included. This shows that with only pre-EMC data
fitted, uncertainties increase by one and a half order of magnitude,
and the increase is exactly the right size to account for the missing
information, i.e. the deviation of results from the true value.

When comparing NNPDFpol1.1 data to the pre-RHIC PDFs,
and post-NNPDFpol1.1 data to NNPDF1.1pol  PDFs, $\chi^2$ values are of
order three-four per datapoint, which means that the
increase of uncertainties in the region of the data not included is
not so dramatic, though still noticeable.
The data not included only provide a relatively minor extension of
kinematic reach, without adding any new process or target.
Also in these cases, the $\chi^2$ value after inclusion of PDF
uncertainties decreases to a value around one, thus showing that
uncertainties are correctly estimated also in near extrapolation.

\begin{table}[t]
\renewcommand*{\arraystretch}{1.60}
\scriptsize
\centering
\begin{tabularx}{.4\textwidth}{|X|ccc|}
\toprule
                 &   NNPDpol1.1 & pre-EMC & pre-RHIC \\
\midrule
$\Delta\Sigma$ & $0.18\pm0.21$       &  $0.83\pm0.74$    & $0.16\pm0.30$   \\
$\Delta g$  & $0.02\pm3.24$       & $1.71\pm4.80$  & $0.95\pm3.87$   \\
\bottomrule
\end{tabularx}

\vspace{0.3cm}
\caption{\small The first moment of the polarized quark singlet and gluon
  distributions, computed using the NNPDFpol PDF sets and the two
  future test pre-EMC and pre-RHIC PDF sets, evaluated at the scale $Q^2=1$~GeV$^2$.}
\label{tab:firstm}
\end{table}
Inspection of the results of  Table~\ref{tab:chi2pol} shows that
$\chi^2$ values
without PDF uncertainties, for datasets which are fitted, is in most cases
smaller, and often rather smaller than one. This is due to fact that,
as already mentioned, statistical and systematic experimental
uncertainties are added in quadrature, because of lack of information
on their correlation. This leads to an overestimate
of the experimental uncertainties, and thus to an underestimate of the
$\chi^2$. Also, as mentioned, correlations between PDF uncertainties instead are
always consistently included, so when a dataset is not fitted and PDF
uncertainties  are dominant the expected
correct result has a $\chi^2$ of order one per datapoint, even if the
$\chi^2$ of the same dataset when fitted is much smaller than one. The
results shown in  Table~\ref{tab:chi2pol} confirm this expectation. 

We  now turn to a comparison of PDFs. Because, as mentioned, only
the three combinations Eqs.~(\ref{eq:trip}-\ref{eq:oct}) of PDFs are
accessible, it is more convenient to look directly at these, and the
gluon. They are compared in Figure~\ref{fig:polpdfs}  for the
 pre-EMC, pre-RHIC PDF and NNPDFpol1.1 sets. The pre-EMC PDFs
are perfectly compatible with
NNPDFpol1.1, within their extremely large uncertainties. The pre-RHIC
and NNPDFpol1.1 polarized quarks are almost identical, while the
pre-RHIC polarized gluon is somewhat more uncertain than the
NNPDFpol1.1 gluon, especially at
large $x\gtrsim0.1$, but in very good agreement with it.

As discussed in Section~\ref{sec:poldat} much of the interest in
polarized PDFs is related to the EMC discovery that the first moment
of the polarized quark distribution is unusually small, and in fact
compatible with zero within uncertainties. This also raised interest
in the value of the first moment of the polarized gluon
distribution, both because of its possible role in carrying a sizable
fraction of the proton spin, and also, in explaining the smallness of
the quark first moment at low scale due to its mixing with it driven
by the axial anomaly~\cite{Altarelli:1988nr}. Note that, as mentioned,
the first moment of the triplet and octet combinations
Eqs.~(\ref{eq:trip}-\ref{eq:oct}) are instead fixed by weak meson
decay constants.

The values of the polarized quark singlet and gluon
first moments are collected in Table~\ref{tab:firstm} for the
NNPDFpol1.1, pre-EMC and pre-RHIC polarized PDF sets. Of course, the
qualitative behavior of the first moments is the same as that of the
PDFs seen in Figure~\ref{fig:polpdfs}: they are all compatible within
uncertainties, with pre-EMC uncertainties much larger, and pre-RHIC uncertainties
somewhat larger for the quark and rather larger for the gluon in
comparison to NNPDFpol1.1. The ``spin crisis'', with the polarized
singlet quark first moment compatible with zero, and significantly
different from a value of order one, is clearly seen already in the
pre-RHIC PDF set. On the other hand, no conclusion on the polarized
quark first moment can be drawn from pre-EMC data: the result is
compatible both with zero and with one, within its large
uncertainty. Even with contemporary fitting methodology, no hint of the ``spin
crisis'' can be seen in pre-EMC data. The spin crisis was a real
experimental surprise.
Interestingly, knowledge of the
gluon first moment has not improved much over the years: both a
vanishing value, or a large value (a possible explanation of the
``spin crisis'') are compatible with both EMC and contemporary data.

Finally, in Figure~\ref{fig:poldata} we compare predictions for selected
pre-EMC and pre-RHIC data, and in Figure~\ref{fig:poldatab} for selected
post-NNPDFpol1.1 data, obtained using the three polarized PDF sets.
The excellent compatibility of the three PDF sets with each other and
with all the NNPDFpol1.1 data is clear, with the uncertainties of the
pre-EMC PDFs very large, and of the right amount to account for the
missing information from the subsequent data. The post-NNPDFpol1.1
data pose more stringent requirements on the NNPDFpol1.1 PDFs, whose
uncertainties are just large enough to accommodate them, as we had
seen already from the $\chi^2$ values of Table~\ref{tab:chi2pol}. It
is apparent from these plots how the PDF uncertainty correctly
accounts for the deviation between data, and predictions obtained
using future test PDFs.

\section{Conclusions and outlook}
\label{sec:conc}

Testing and validating the generalization power of machine learning
tools is one of the most difficult and important problems of
artificial intelligence. In the context of PDF determination, this is
is an especially delicate issue since it is the generalization of PDFs
outside the data region that makes it possible to obtain physics
predictions at hadron colliders with finite uncertainties.
This issue manifests itself as the problem of forward-backward compatibility
of PDFs. Indeed,  when PDFs are
assumed to have some fixed underlying functional form, it is a not
uncommon occurrence that subsequent data require enlarging the
functional form, with the paradoxical consequence that adding new
data results in larger, rather than smaller uncertainties (see
e.e. Ref.~\cite{Hou:2019efy}). PDFs determined using NNPDF methodology
so far have been free of this problem, with later PDF sets always
compatible within uncertainties with previous ones.

Here, we have suggested the use of forward-backward compatibility
as a criterion for validating a
PDF methodology. Using forward compatibility as a criterion for
validation of a current methodology is actually more stringent than
simply checking the forward compatibility of existing PDF sets. This
is because as the methodology improves, methodological uncertainties
become smaller with fixed underlying data, as a consequence of the fact
that more recent methodologies are more efficient in extracting
information from the data.

We have seen that the current most recent
NNPDF methodologies in the unpolarized and polarized case pass the
future test, thereby showing that the NNPDF methodology correctly
extrapolates outside the data region. It is interesting to ask
whether the method can also be used for a comparative assessment of
different PDF fitting methodology, possibly not only to
validate a posteriori the extrapolation methodology, but also to
actually optimize it. This question will be left to future investigations.

\section*{Acknowledgments}
We thank all the members of the NNPDF collaboration and the N$^3$PDF team for
innumerable discussions on PDF determination and validation, and
especially Christopher Schwan for a critical reading of the manuscript
and many insightful comments and Juan Rojo for several useful
suggestions on the draft. S.F. is
very grateful to Micha\l\ Praszalowicz for managing to organize the LX
Zakopane school as an online meeting in the current difficult
circumstances, and to Marek Karliner for suggesting during the meeting
to investigate the proton spin crisis using future tests. 
This work is  supported by the European Research Council under
the European Union's Horizon 2020 research and innovation Programme
(grant agreement n.740006). ERN is supported by the UK STFC grant ST/T000600/1.

\bibliographystyle{UTPstyle}
\bibliography{future}

\end{document}